\documentclass[prd,floatfix,nofootinbib,showkeys]{revtex4-2}
\usepackage[utf8]{inputenc}
\usepackage[T1]{fontenc}
\usepackage{amsmath}
\usepackage{amsfonts}
\usepackage{amssymb}
\usepackage{graphicx,epstopdf}
\usepackage{datetime}
\usepackage{hyperref}

\graphicspath{{Figures/}}

\begin{document}	
	\title{Lepton-induced reactions on nuclei in a wide kinematical regime}
	
	\author{U. Mosel}
	\email[Contact e-mail: ]{mosel@physik.uni-giessen.de}
	\author{K. Gallmeister}
	\affiliation{Institut f\"ur Theoretische Physik, Universit\"at Giessen, 35392 Giessen, Germany}
\date{\today}

\begin{abstract}
Inclusive differential cross sections for various $e A$ and $\nu A$ reactions are analyzed within the GiBUU theoretical framework and code. The treatment of electron-nucleus reactions has been improved by implementing a parametrized description of electron-nucleon interactions for a nucleon. Using the momentum of a nucleon inside the Fermi sea the electron-nucleon cross sections are then Lorentz-boosted to obtain the electron structure functions for nuclei. The neutrino structure functions are obtained from the ones for electrons by a transformation that involves the axial formfactors and kinematical factors that account for the difference of vector and axial currents. Special emphasis is put on analyzing data from various different experiments in different neutrino energy regimes with one and the same theoretical input. Good agreement is reached for a wide kinematical regime, from MicroBooNE to the medium energy MINERvA experiment.
\end{abstract}


\maketitle
\newpage

\section{Introduction}

Interactions of photons and electrons with nuclei have given us interesting information on the electromagnetic vector couplings of bound nucleons and on in-medium properties of hadrons \cite{Drechsel:1989ab,Boffi:1996ikg}. The interactions of neutrinos with nuclei can supplement this information by yielding information on in-medium axial couplings \cite{Gallagher:2011zza}. 

There is also a very practical side of such studies: the determination of neutrino mixing parameters, neutrino mass ordering and a possible CP violating phase in long-baseline experiments requires the knowledge of the incoming neutrino energy on an event-by-event basis \cite{Diwan:2016gmz,Katori:2016yel,Mosel:2016cwa}. Different from all other nuclear or high-energy experiments this neutrino energy is not known. It has to be reconstructed from observations of the final state of the reaction. This reconstruction requires event generators \cite{Mosel:2019vhx,NeutrinoGen:2022} that can follow the full time-development of the reaction from the initial reaction of the incoming neutrino with a bound and Fermi-moving nucleon to the final state of the reaction with outgoing nucleons and mesons. It is obvious that this reconstruction is the better the more reliable these generators are \cite{Mosel:2016cwa}. 

The reliability of a generator requires that the generator is built on state-of-the-art nuclear physics. The generator thus has to be consistent in its description of the various different subprocesses, such as, e.g., pion production and absorption, quasielastic scattering, etc. This requirement ensures that there are no redundant, unphysical degres of freedom to tune; tuning is permissible only within the theoretical uncertainties of the description of these processes. Furthermore, the generator results have to be confronted with experimental data. Observations of discrepancies with experimental data from different experiments can help to improve the underlying theory and its practical implementation. 

So far, experimental analyses of data from long- or short-baseline experiments have often followed another approach. They have tuned their favorite generator, often built on a patchwork of different, sometimes inconsistent theories for the various subprocesses, to their data \cite{Bronner:2016gmz,Tena-Vidal:2021rpu,GENIE:2021wox,MicroBooNE:2021ccs,NOvA:2020rbg}. Special tunes were thus obtained for the different experiments, covering a wide range of neutrino energies. This usual way of tuning generators is dangerous since the choice made on which parameters to tune, in particular if they are superficially redundant, may hide the correct physics \cite{Mosel:2019vhx}. An example is the original finding by the MiniBooNE experiment of a significantly increased axial mass \cite{MiniBooNE:2007iti} which later on turned out to be a consequence of the missing 2p2h reaction component \cite{Martini:2009uj,Nieves:2011pp}. Another potential problem of this 'tuning approach' is connected with the fact that the very same generator that is later on tuned to reproduce the data is already used during the data-analysis (without the final tune) to determine the experimental efficiencies. Such an approach would need in principle an iteration of data-analysis and data tuning that is hardly ever done. 

What is still missing up today is an attempt to tune a generator simultaneously to all experiments. It is, therefore the aim of the present paper to try to analyze inclusive cross-section data from fairly low-energy experiments, such as MicroBooNE, to higher-energy experiments, such as the MINERvA experiment at a middle energy. In this study we will point out the dominant reaction processes in these different energy regimes.
 
 Another aim of the present paper is to present the theoretical foundations of new implementations of resonance and background contributions, both in electron- and neutrino-induced reactions, in the new version 2023 of the GiBUU theory and generator \cite{gibuu}. Together with Refs.\ \cite{Leitner:2009zz, Buss:2011mx,Gallmeister:2016dnq}  all the theoretical implementations are then well documented; the code is available from \cite{gibuu}.

\section{Inclusive cross sections}

For the reconstruction of the incoming neutrino energy from the outgoing particles the full final state is needed and inclusive cross sections in which only the outgoing lepton is observed are not enough. They do, however, present a necessary check. Therefore, in the present paper we investigate if a common description of different neutrino-induced inclusive measurements on nuclei can be achieved within the GiBUU theory framework and code \cite{Buss:2011mx,gibuu}. For this comparison we use available double-differential cross sections from different experiments. We then try to identify the reaction processes that are most important in any given experiment and attempt to optimize their description within the theoretical uncertainties. A study with a similar aim in mind, based on the SuperScaling model and state-of-the-art implementations of nucleons resonance excitations, has very recently been published \cite{Gonzalez-Rosa:2023aim}. 

The double-differential inclusive cross section for events with an incoming neutrino energy $E$ is given by
\begin{eqnarray} \label{eq:nusigma}
	\frac{{\rm d}^2\sigma}{{\rm d}E' {\rm d}\Omega} (E) &=&  \frac{G^2}{2\pi^2} \left(\frac{M_W^2}{M_W^2 + Q^2}\right)^2\,E'^2\,\left[2 W_1(Q^2,\omega) \sin^2(\frac{1}{2}\theta) + W_2(Q^2,\omega) \cos^2( \frac{1}{2}\theta) \right. \nonumber \\
	 & & {} \left. \mp \, W_3(Q^2,\omega) \frac{E + E'}{m}\sin^2(\frac{1}{2}\theta) \right] ~.
\end{eqnarray}
Here $E' = E - \omega$ is the energy of the outgoing lepton and $\Omega$ is its solid angle; $m$ is the nucleon mass. The three structure functions are all functions of the four-momentum squared $Q^2$ and of the energy transfer $\omega$: $ W_i = W_i(Q^2,\omega)$. The structure functions $W_1$ and $W_2$ contain the incoherent sum of the vector and axial-vector current contributions. The third structure function $W_3$ is due to the interference of vector and axial currents. Its sign depends on whether the incoming lepton is a neutrino ('-' sign) or an antineutrino ('+' sign).

In accelerator-based neutrino experiments the incoming beam energy is not sharp, but instead described by an energy-distribution, called flux, $\Phi(E)$, so that the measured cross section is given by
the flux average
\begin {equation}
\langle	\frac{{\rm d}\sigma}{{\rm d}E' {\rm d}\Omega}\rangle = \int \Phi(E)\, \frac{{\rm d}\sigma}{{\rm d}E' {\rm d}\Omega} (E) \, {\rm d}E ~;
\end{equation}
here the flux is assumed to be normalized to 1. The flux decouples from the structure functions. The latter are thus identical for different experiments using the same target, but different flux distributions. They contain all the nuclear physics information. Ideally, any tuning should thus focus on tuning the structure functions, and not the cross sections which are different in different experiments even when they are using the same target.

Experiments in different energy regimes cover different, but often overlapping, regions in the $(Q^2,\omega)$ plane because they are sensitive to different elementary processes. For example, the MicroBooNE and T2K experiments work at such a low neutrino energy that essentially only QE scattering, $\Delta$ resonance excitation with its conected pion production and 2p2h excitations play a role; both the momentum transfer $Q^2$ and the energy transfer $\omega$ are fairly small. On the contrary, the inclusive cross section at the MINERvA medium energy (ME) experiment with its average energy of 6 GeV (and also at DUNE)  receives a large contribution from higher lying resonances, where $\omega$ is large, and Deep Inelastic Scattering (DIS) where both $Q^2$ and $\omega$ are large. That these experiments work in different energy regimes is thus helpful in determining the structure functions in as wide a kinematical range as possible. These structure functions are built from contributions from different processes (such as QE scattering, 2p2h excitations, resonance excitations etc). Therefore, a unified description of many of such experiments with one and the same physics input and parameter set can also give essential information on the underlying reaction mechanisms which prevail in different kinematical regions. For the generators, widely used by experimenters, such a consistent tune has not emerged yet. For example, GENIE has been used to obtain quite different tunes to data in the GeV region (for MINERvA \cite{MINERvA:2020zzv,MINERvA:2021owq} and NOvA \cite{NOvA:2020rbg}) and the sub-GeV region (MicroBooNE \cite{MicroBooNE:2021ccs}). A simultaneous description of all of these data, together with those from T2K, is thus still a challenge. Another problem with the studies just mentioned is that even within one experiment different tunes (versions) of one and the same generator are sometimes used without giving details about the physics changes made by going from one to the other.

\section{Model}
The basis of our investigation is the GiBUU theory framework. Its theoretical (and practical) ingredients are described in some detail in \cite{Leitner:2009zz,Buss:2011mx,Gallmeister:2016dnq} and its source code is freely available from \cite{gibuu}. The code has been widely used in the description of hadron-nucleus \cite{Gallmeister:2009ht}, nucleus-nucleus \cite{Larionov:2020fnu} and electron- and photon-nucleus reactions \cite{Lehr:2003ht,Buss:2006vh}. It can be used as a generator for the complete final state, which is obtained from quantum-kinetic transport theory \cite{Kad-Baym:1962,Buss:2011mx}. It can also be used for a calculation of inclusive cross sections. In this latter case no time-development of the reaction is performed, but the calculation of the initial cross sections takes the final state potentials into account.

Here we now just mention the ingredients which are most important for the following discussions:
\begin{itemize}

\item Unlike in other generators in GiBUU the target nucleons are bound in a momentum-dependent mean-field potential $U(\bf{r},\bf{p})$ with their momenta given by a local Thomas-Fermi distribution. The same potential $U$ is seen by the outgoing particles. This necessitates a numerical integration of the trajectories of outgoing particles. It also complicates energy-momentum conservation in final-state collisions which usually requires some numerical iterations in $E$ and $\mathbf{p}$ \cite{Buss:2011mx}. 

\item The quasielastic (QE) scattering is modelled within the impulse approximation with an axial dipole form factor with an axial mass of 1.0 GeV. There is no additional RPA correction, as often used in other generators such as GENIE. It has been shown that these correlations are significantly diminished if nucleons are bound in a mean-field potential \cite{Pandey:2016jju,Nieves:2017lij} and energy transfers are larger than about 40 MeV.

\item The meson exchange current (MEC) component is obtained from a fit to electron data \cite{Christy:2015,Bodek:2022gli} \footnote{The present version uses the parametrization given in \cite{Bodek:2022gli} with parameters corrected through private communication with E. Christy. The parametrization of \cite{Christy:2015} was used in earlier versions of GiBUU. It is still available in the code and can be chosen by a parameter in the job card.}.  The new parametrization consists of two Gaussians, one centered below the QE peak and another one centered in the dip region between QE and $\Delta$ peak (see Appendix). The former one adds transverse strength at the QE peak; it goes together with a quenching of the longitudinal cross section in this region \cite{Bodek:2022gli} so that the total QE peak is hardly affected. 

\item The nucleon resonance region (defined to reach up to an invariant mass of $W =2$ GeV) is described by implementing  nucleon resonances in this mass range, together with their decay channels and background contributions \cite{Buss:2011mx}. In earlier work (up to version 2021)  we took the vector formfactors from the MAID 2007 analysis of electron scattering data. MAID is based on an analysis of $\pi\,N$ production and thus does not provide information about the $2\pi\,N$ and higher reaction channels. Thus, the background contributions beyond the $\Delta$ resonance had to be inserted by hand. 

Therefore, for electron scattering we have now switched the description of this mass region to the analysis of Bosted and Christy \cite{Christy:2007ve,Bosted:2007xd} which has obtained fits to proton and neutron data up to a mass of $W=3$ GeV and up to momentum transfers of $Q^2=8$ GeV$^2$; the fits contain both the resonance and the background contributions. For neutrinos we still calculate the resonance contributions as outlined in detail in \cite{Leitner:2009zz}. The vector couplings are still taken from MAID2007; the axial couplings are obtained from PCAC with a $g_A = 1.17$; the formfactors are given by a dipole form with axial mass of 1.0 GeV. A special role is played by the $\Delta$ resonance \cite{Leitner:2006ww,Lalakulich:2010ss} which is reasonably constrained by pion production data and where a somewhat more complicated formfactor is chosen \cite{Leitner:2006ww}. Still, there is some ambiguity in its different axial formfactors and its coupling constants \cite{Lalakulich:2006sw} which we will exploit later on in this paper. We assume that in this resonance region the decay products of the background contribution are dominated by 1$\pi$ and $2\pi$ events.

\item Both in the resonance region as well as in the transition region between $W=2$ and $W = 3$ GeV also so-called background terms contribute which are connected with either 1-pion t-channel production or even 2- and higher-pion and other meson $t$-channel production processes. For $W > 2$ GeV these terms also contain contributions from higher-lying, broad resonances. For electrons we now obtain the background cross sections from the same Bosted-Christy analysis \cite{Christy:2007ve,Bosted:2007xd} as for the resonances.  The electron observables are then converted to those for the neutrinos by using the same transformation as for the 2p2h contribution (for details see Sect. \ref{s:neutrinos}).

\item Events in the mass region above $W$ = 2 GeV are called 'DIS'. For the transition region between 2 and 3 GeV we still use the Bodek-Christy parametrization for the cross section, the wording 'true DIS' is used for processes with $W$ > 3 GeV. The decay products of the background contribution are provided by string fragmentation as implemented in PYTHIA \cite{Sjostrand:2006za}; this latter assumption is not important for the inclusive cross sections discussed in this present paper.

For invariant masses above 3 GeV we rely on PYTHIA for a description of the deep inelastic scattering (DIS) events both for its cross section and for the actual event generation. Contrary to our earlier attempt to cut out the low-$Q^2$ events \cite{Gallmeister:2016dnq} we now no longer employ such a low-$Q^2$ correction. Only without such a correction we reproduce the measured high-energy neutrino-nucleon cross sections \cite{Lalakulich:2012gm}.

\end{itemize} 

In the following section we give the relevant details on the actual implementation of the points just mentioned.

\subsection{Nuclear Groundstate}
The groundstate of the target nucleus is prepared by first choosing a realistic density distribution. Using a Skyrme-type energy-density functional we obtain a single particle potential $U(\mathbf{r},\mathbf{p})$ that depends on position $\mathbf{r}$ and momentum $\mathbf{p}$. The momentum-distribution is approximated by that of a local Fermigas. For simplicity we then transform this potential into a Lorentz scalar $U_S$ \cite{Buss:2011mx} so that the hole spectral function in the nuclear groundstate becomes in the quasiparticle approximation (off-shell effects neglected)
\begin{equation}    \label{eq:SF}
	\mathcal{P}_h(\mathbf{p},E) = g \int_{\rm nucleus} {\rm d}^3r \, f(\mathbf{r},\mathbf{p},t=0) \, \delta\left(E  - \sqrt{\mathbf{p}^2 + {m^*}^2(\mathbf{r},\mathbf{p})} + m \right)~.
\end{equation}
Here $f(\mathbf{r},\mathbf{p},t=0)$ is the single-particle Wigner transform of the one-body density matrix at time $t = 0$. In the local Thomas-Fermi approximation it is given by $ \Theta(p_F(\mathbf{r}) - |\mathbf{p}|)$. Here $p_F(\mathbf{r}) \propto \rho^{1/3}(\mathbf{r})$ is the local Fermi momentum, $m$ is the free nucleon mass and  $m^* = m - U_S$ is the effective mass that contains the potential $U_S(\bf{r},\bf{p})$; $g$ describes the spin-isospin degeneracy. 

Because of the integration over the nuclear volume in Eq.\ (\ref{eq:SF}) the spectral function is no longer a simple $\delta$-function in the $(\mathbf{p},E)$ plane. It is instead smeared out, without any shell effects due to its semi-classical nature \cite{Alberico:1997jg}. While spectral functions obtained from nuclear many body theory show signs of a shell structure (see, e.g., Fig.\ 4 in \cite{Sobczyk:2023mey}) this structure is being averaged out in neutrino experiments where the energy-transfer cannot be measured.

The momentum-dependence of $U$ is obtained by fitting the Welke parametrization \cite{Gale:1989dm} to (p,A) data \cite{Cooper:1993nx}. When exclusive or semi-inclusive reactions are treated the same potential $U$ is then used for the propagation of  final state nucleons. This is in contrast to the treatment of spectral functions in other generators where the potential is hidden in the spectral function and cannot be used for the final state propagation.

Technically, the spectral function is represented by an integral over the  sum over many 'testparticles'
\begin{equation}
	f(\mathbf{r},\mathbf{p},t) \propto \sum_{j=1}^N \delta(\mathbf{r} - \mathbf{r}_j(t))\, \delta(\mathbf{p} - \mathbf{p}_j(t))~;
\end{equation}
see Sect.\ 2.4.3 in \cite{Buss:2011mx}. Here $\mathbf{r}_j(t)$ and $\mathbf{p}_j(t)$ are the position and momentum of the $N$ testparticles that make up the nucleon phase-space distribution.

\subsection{Electron Cross Sections}
The cross section for inclusive electron scattering is given by
\begin{eqnarray} \label{eq:esigma}
	\frac{{\rm d}^2\sigma}{{\rm d}E' {\rm d}\Omega} (E) &=&  \frac{4\alpha^2}{Q^4} \,E'^2\,\left[2 W_1^e(Q^2,\omega) \sin^2(\frac{1}{2}\theta) + W_2^e(Q^2,\omega) \cos^2( \frac{1}{2}\theta) \right]
\end{eqnarray}	

\begin{figure}
	\centering
	\includegraphics[width=0.7\linewidth]{"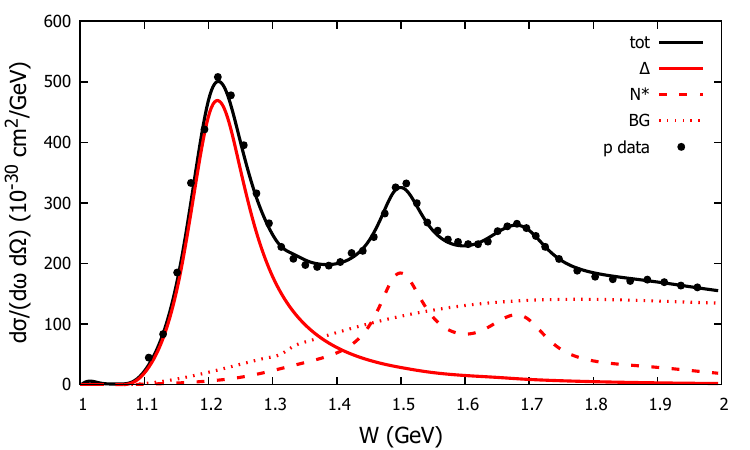"}
	\includegraphics[width=0.7\linewidth]{"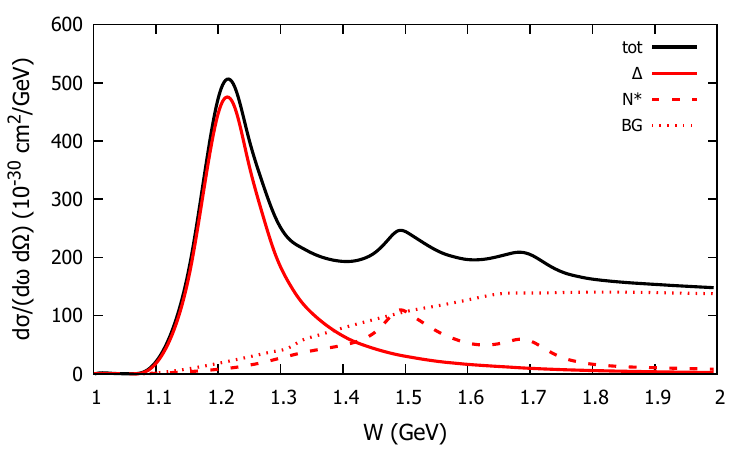"}
	\caption{Inclusive cross section for electron scattering on a proton (top) and a neutron (bottom) in the resonance region as a function of the invariant mass $W$ at an electron energy of 2.238 GeV and an angle of 21.95 degrees. The various contributions to the total cross section are indicated in the figure; BG denotes the non-resonant background contribution. Data are from \cite{JeffersonLabHallCE94-110:2004nsn,JLABdata_E94-110}.}
	\label{fig:ep_2.238_21.95}
\end{figure}

The two electron structure functions $W_i^e$ have been fitted to proton and neutron (through Deuterium) data by the authors of Ref.\ \cite{Christy:2007ve,Bosted:2007xd}. An example for the quality of this fit is shown in Fig.\ \ref{fig:ep_2.238_21.95}
where we show the cross section for electron scattering on the proton and on a neutron. This figure just illustrates how well the Christy-Bosted parametrization \cite{Bosted:2007xd,Christy:2007ve} works over the resonance region. We note here that - based on this success - this parametrization has also been used in calculations within the scaling model \cite{Maieron:2009an,Megias:2016lke,Gonzalez-Rosa:2022ltp}.

To obtain the electron structure functions for nuclear targets we now take the target nucleons with their given momentum distribution (see Eq.\ (\ref{eq:SF})) and Lorentz-transform to the restframe of each testparticle. This Lorentz-transform takes the binding energy of the testparticle (nucleon) into account. We then evaluate the cross section in this restframe using the expressions described above. After that we Lorentz-transform the cross section back to the nuclear (lab) restframe \cite{Byck:1973}
\begin{equation}
	\left(\frac{1}{p'}\,\frac{{\rm d}^2\sigma}{{\rm d}E' {\rm d}\Omega} \right)_{\rm lab frame}  = 	\left(\frac{1}{p'}\,\frac{{\rm d}^2\sigma}{{\rm d}E' {\rm d}\Omega}\right)_{\rm nucleon rest frame} ~.
\end{equation}
The transformation just described is performed both for the resonance contribution and the background contribution. 

The Lorentz-transformation from the lab to the nucleon restframe does not change $Q^2$, but it does change the energy transfer $\omega$ and, correspondingly, also the three-momentum transfer. The overall effect is thus different from the one usually used \cite{Benhar:2006wy} where $\omega$ is changed but the three-momentum transfer is kept the same, thus changing also $Q^2$. Essential for our treatment is the assumption that the cross section on a bound nucleon at rest is the same as for a free nucleon at rest. In Ref.\ \cite{Lalakulich:2012gm} we have shown that the binding energy has only a minor influence on the cross sections.

\subsection{Neutrinos}  \label{s:neutrinos}

For electrons the two structure functions $W_{1,2}^e$ are well defined, based on the fit of Bosted and Christy \cite{Christy:2007ve,Bosted:2007xd} to the proton and neutron cross sections. For neutrinos  these are only very roughly known since data on elementary targets are very rare \cite{Formaggio:2012cpf}. Furthermore, the structure function $W_3$, which originates in the interference of vector and axial coupling, appears only for neutrinos and thus is not known. For neutrinos we, therefore, retain our original description for quasielastic \cite{Gallmeister:2016dnq} and resonance contributions described in detail in \cite{Leitner:2006ww,Leitner:2006sp,Leitner:2009zz}; in addition the DIS contribution is obtained from PYTHIA \cite{Sjostrand:2006za}. 

This leaves the neutrino structure functions for the background and the 2p2h contributions to be determined. Both are dominantly transverse \cite{Christy:2007ve,Martinez-Consentino:2021vcs}. Other models \cite{Bodek:2010km,Gonzalez-Rosa:2022ltp} rely on approximate uses of a parton model to determine these neutrino structure functions. While this is the appropriate description for very high neutrino energies it is not clear how well these approximations work at the energies of a few GeV relevant for present-day and planned neutrino experiments. We, therefore, follow the alternative course to use relations originally developed for low-energy nuclear excitations.

 Walecka has shown that in a model of independent particles the transverse electron structure function can be related to the neutrino structure function by the relations \cite{Walecka:1975,OConnell:1972edu}
\begin{eqnarray}    \label{W1nu}
	W_1^\nu &=& \left[1 + \left(\frac{2 m}{\mathbf{q}}\right)^2 \left(\frac{G_A(Q^2)}{G_M(Q^2)}\right)^2\right] \,2 (\mathcal{T} + 1)\, W_1^e \nonumber \\
	W_3 &=& 2 \left(\frac{2 m}{\mathbf{q}}\right)^2 \frac{G_A(Q^2)}{G_M(Q^2)}\, 2 (\mathcal{T} + 1)\, W_1^e	~.
\end{eqnarray}
Here $m$ is the nucleon mass, $\mathbf{q}$ is the 3-momentum transfer, $G_A$ is the axial coupling constant, here taken to be $G_A(0) = -1.23$ and $G_M$ is the electromagnetic isovector coupling constant $G_M(0) = 4.71$. Both coupling constants effectively depend on $Q^2$ through the standard dipole formfactors with vector and axial masses equal to 0.84 and 1.03, respectively. Finally, $\mathcal{T}$ is the isospin of the nucleus. While Walecka  used this relation only for the description of low-lying states in nuclei \cite{Walecka:1975}, it is also valid for the description of high-lying states provided that they can be described in a single particle model. The latter assumption is often invoked in neutrino-nucleus interactions, e.g.\ by using the impulse approximation for QE scattering. 

In general $W_1^\nu$ is given by the sum of squares of the vector and the axial-vector current contributions whereas $W_3$ is given by the interference term of both. This is reflected in the structure of the expressions given in (\ref{W1nu}). $W_1^e$ contains the electromagnetic coupling constant $G_M^2$ which, in the second term of $W_1^\nu$, is divided out and replaced by the axial coupling constant $G_A^2$.  Similarly, $W_3$, being due to the vector-axialvector interference, is just given by replacing one factor $G_M$ in $W_1^e$ by the axial coupling constant $G_A$. The kinematical factor $(2m)/\mathbf{q}$ appearing together with the axial contribution gives the connection between the vector and the axial-vector single particle matrix element. The structure of the connecting expression of eq.\ (\ref{W1nu}) is thus quite natural and appealing. Since the single-particle nature of high-lying excitations underlying eq.\ (\ref{W1nu}) is an approximation, comparisons with experiment have to show if it works in the region of interest here. 

We exploit this general structure by using these relations not only for the pion background but also for the 2p2h contribution.  Even though the latter is not a single-particle process, we have already shown in Ref.\ \cite{Gallmeister:2016dnq} that this connection between the electron and the neutrino structure functions works very well for the 2p2h processes both for neutrinos and for antineutrinos. This is a non-trivial test since together with the structure function $W_1$ also the structure function $W_3$ is obtained without any further adjustment.   A very similar result has recently also been obtained by the authors of Ref.\ \cite{Martinez-Consentino:2021vcs}.

The quantity $\mathcal{T}$ in Eq.\ (\ref{W1nu}) is the isospin of the target nucleus. It arises from using the Wigner-Eckart theorem to relate the isovector neutrino process to the corresponding electron interaction by assuming that isobaric analogue states are involved in both reactions. Since this is uncertain we treat $\mathcal{T}$ as an integer parameter, with the parameter adjusted with the same value for the different kinematical regions.  

\section{Results}

\subsection{Electrons}
In this section we show some results obtained with the new implementation of ($e$,nucleon) cross sections for ($e$,A) reactions at a fixed beam energy and scattering angle. 

\begin{figure}
	\centering
	\includegraphics[width=0.7\linewidth]{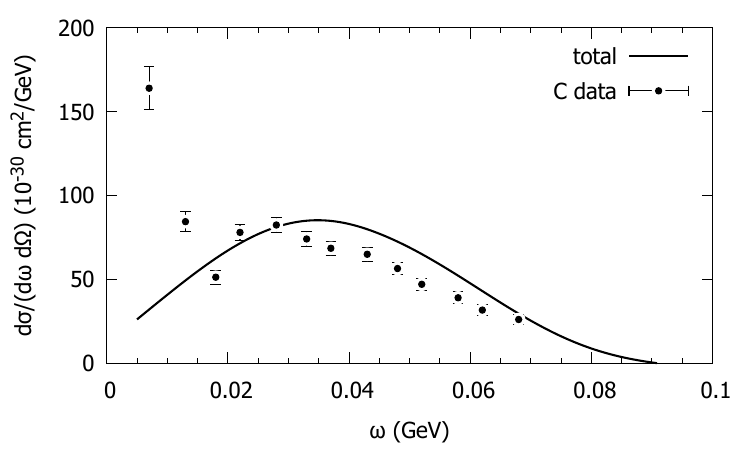}
	\caption{Inclusive cross section for the (e,C) reaction at 0.2 GeV, 60 degrees as a function of energy transfer $\omega$. The curve gives the GiBUU result for QE scattering only. All other contributions are negligeably small. The data are from \cite{Benhar:2006er}.}
	\label{fig:eC_0.2_60}
\end{figure}

A first example for a low energy is given in Fig.\ \ref{fig:eC_0.2_60}. The figure shows the double-differential cross section at the fairly low beam energy of 0.2 GeV as a function of energy transfer $\omega$. 
At this low energy only Quasielastic Scattering (QE) plays a role. The shape of the QE peak is reasonably well described; on the high-$\omega$ side of the peak the discrepancy is only a slight shift of about 5 MeV in the energy transfer. An accuracy of that order is outside the scope of any semi-classical reaction theory. The strong disagreement at low masses is due to nuclear structure effects. GiBUU works with a semi-classical description of the nuclear ground state and thus describes nuclear excitations in the target only on average. A quantitative description of these excitations requires a Continuum RPA calculation such as those already studied about 30 years ago in \cite{Kolbe:1994xb,Kolbe:1995af} and, more recently, in \cite{Pandey:2014tza,Jachowicz:2019eul}. In the present example the energy transfer at the QE peak is about 35 MeV; this clearly defines a lower limit of the applicability of GiBUU because a quasiclassical description cannot account for any specific nuclear target excitations. The code will technically run also at lower energy transfers, but the physics implemented is no longer sufficient for these lower energies.

\begin{figure}
	\centering
	\includegraphics[angle = -90,width=0.7\linewidth]{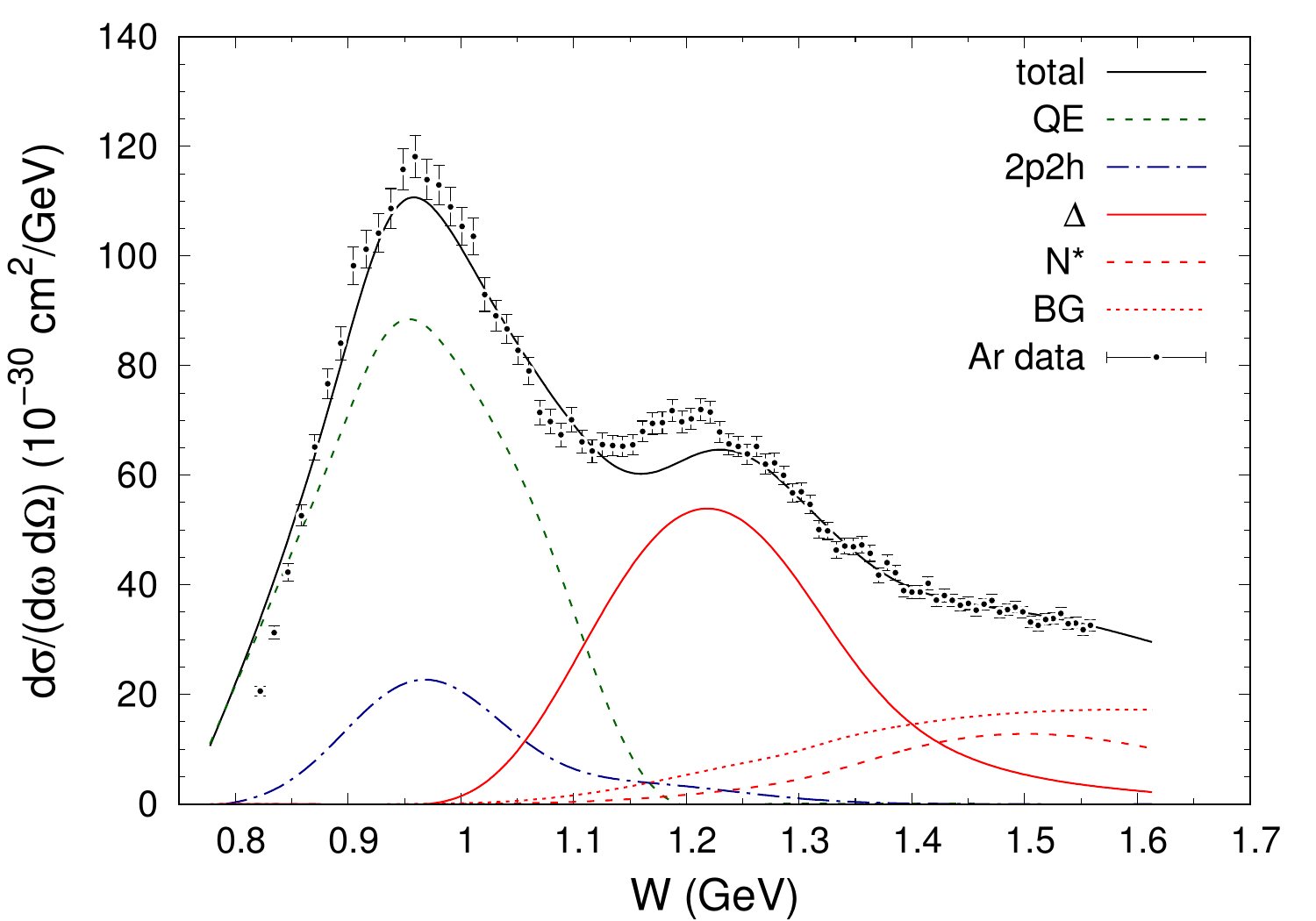}
	\caption{Inclusive cross section for the (e,Ar) reaction at 2.222 GeV, 15.541 degrees as a function of invariant mass. The various curves give the contributions of specific first-interaction processes to the total cross section. The BG contribution contains the total background contribution. The data are from \cite{JeffersonLabHallA:2018zyx}.}
	\label{fig:eAr}
\end{figure}

Results for the scattering on an Ar target at kinematics close to the ones shown in Fig.\ \ref{fig:ep_2.238_21.95} for the elementary reaction are given in Fig.\ \ref{fig:eAr}. 
The abscissa in this plot, the invariant mass $W$, is taken to be $W^2 = m^2 + 2 m \omega - Q^2$; here $m = 938$ MeV is the free mass of the nucleon, $\omega$ the energy transfer and $Q^2$ the squared four-momentum transfer. The invariant mass $W$, as defined here, does not include any effects of nuclear binding or of Fermi-motion. 

The agreement of the calculated result with data is good and better than that obtained with the earlier version (v2019) of GiBUU which was shown in \cite{Mosel:2018qmv}. We attribute this better agreement in the QE-peak region to the use of the new parametrization for the MEC component obtained by Bodek and Christy \cite{Bodek:2022gli}. The obvious disagreement around $W \approx 1.2$ GeV might be due to a too small high-mass component in the MEC parametrization (see the appendix). The resonance region beyond the $\Delta$ resonance, where higher resonances and background contribute, is perfectly well described. 

The overall agreement is significantly better than that shown by the GENIE generator results obtained both with an implementation of the GSUSAv2 scaling result as well as with the G2018 version (see fig.\ 10 in \cite{PhysRevD.103.113003} ), the latter being a tuned version of GENIE v3, which both overshoot the cross section for higher $W$ starting at the $\Delta$ resonance.

\begin{figure}
	\centering
	\includegraphics[width=0.7\linewidth]{"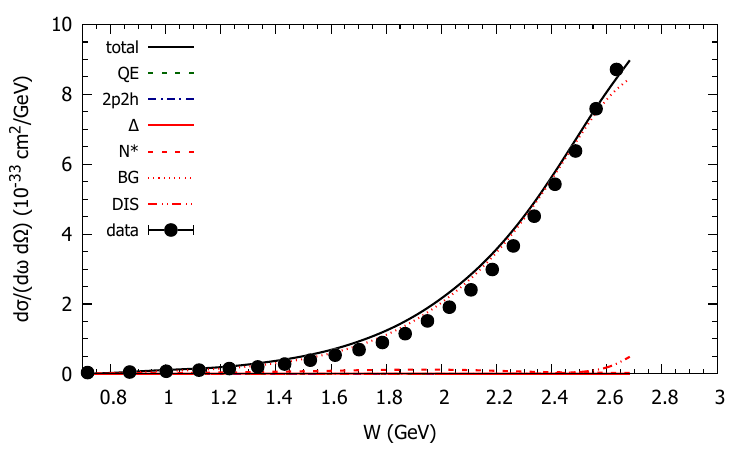"}
	\caption{Inclusive double-differential cross section for scattering of an electron off a Carbon target at a beam energy of 5.766 GeV and a scattering angle of 50 degrees. Data are from \cite{Benhar:2006er}}.
	\label{fig:ec5}
\end{figure}

In order to check the SIS region above about $W = 2$ GeV we show as a last example in Fig.\ \ref{fig:ec5} a reaction at a higher electron energy. 
The cross section here is dominated by the background amplitude, only at the highest $W$ some DIS contribution appears. The cross section is slightly above the data, but the overall agreement again is very good.

\subsection{Neutrino Cross Sections}

Experiments such as MiniBooNE, MicroBooNE and T2K are dominated by a low-energy region where QE scattering and the $\Delta$ resonance play a role. These regions are hardly affected by the recent codechanges in GiBUU v2023. With the previous version of GiBUU we have performed detailed comparisons with T2K results, both for inclusive \cite{Gallmeister:2016dnq} and quasielastic-like events \cite{Mosel:2017anp}. Since these events are dominated by QE scattering, $\Delta$ and the MEC excitation, all of which have not undergone significant changes to the present version, we just refer the interested reader to these just cited papers. On the other hand,  DUNE will be operating in a region where higher-lying resonances and DIS are dominant, i.e. in regions possibly more affected by the recent code improvements.

\subsection{NOvA inclusive data}
We start with the NOvA experiment as an 'in-between' experiment that is sensitive to both the (QE,$\Delta$) and the higher resonance regions; comparisons with results obtained with v2019 of GiBUU were presented in Fig.\ 9 in \cite{NOvA:2021eqi}. The comparisons given there show that this previous version of GiBUU describes the cross-section very well at the most forward angular bin $0.99 < \cos(\theta) < 1.0$ but underestimates the measured inclusive double-differential cross sections at non-forward angles with this underestimation being largest at $0.8 < \cos(\theta) < 0.85$. That the disagreement is angle-dependent hints at an incorrect $Q^2$ dependendence of the cross section on top of a constant tuneable strength factor in the inclusive cross-section. We now try to explore some channels contributing to the inclusive cross section with the help of these NOvA data. We stress that in these explorations we always stay within the theoretical uncertainties of input parameters.

\begin{figure}[h]
	\centering
	\includegraphics[width=0.8\linewidth]{"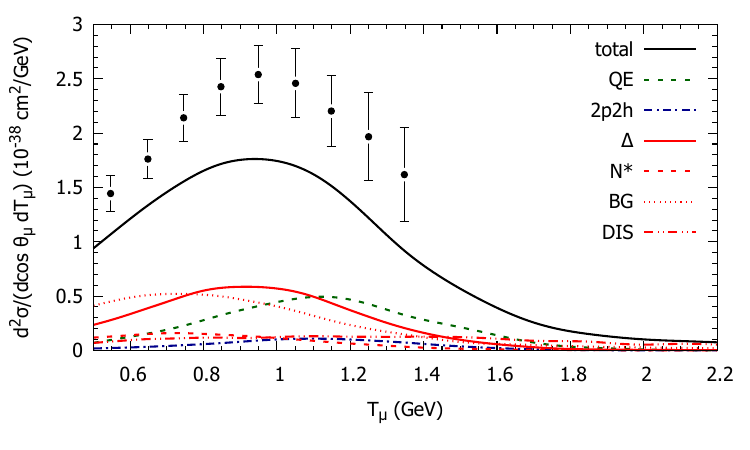"}
	\caption{Double-differential cross sections per nucleon from the NOvA experiment for the angular bin:  $0.8 < \cos \theta < 0.85$. Data are taken from \cite{Cremonesi:2020,NOvA:2021eqi}. The various curves give the contributions of specific first-interaction processes to the total cross section calculated with the 2019 version of GiBUU. The BG contribution contains the total background contribution.}
	\label{fig:nova_2019}
\end{figure}
In order to clarify which particular processes could be responsible we show in Fig.\ \ref{fig:nova_2019} the particular angular bin $0.8 < \cos \theta < 0.85$ \cite{Cremonesi:2020} as obtained in the 2019 version. The figure contains the various contributions to the inclusive cross-section. It is noticeable that the shape of cross section resembles that of only three contributions, those of QE scattering, of $\Delta$ resonance excitation and of the background, whereas all the other contributions are in their sum (about 5- 10\% of the total) flat over the full range of the outgoing muons kinetic energy $T_\mu$. The QE strength is reasonably well determined by the data on QE scattering on the nucleon.

\begin{figure}
	\centering
	\includegraphics[width=0.7\linewidth]{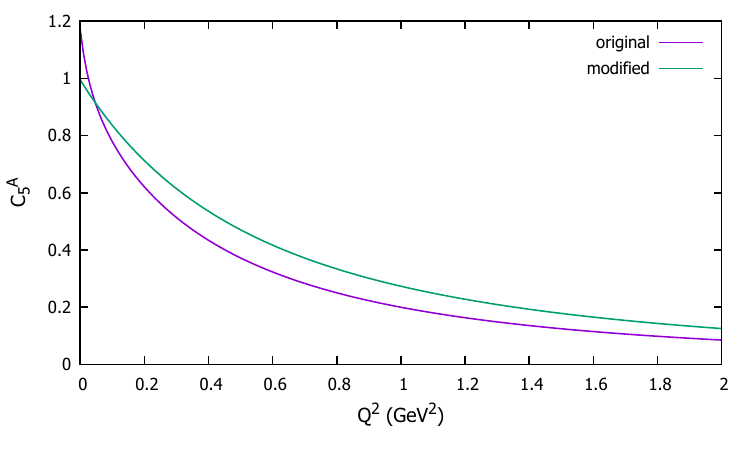}
	\caption[]{Axial formfactor $C_5$ of the $\Delta$ resonance. The curve labelled 'original' is obtained from Eq.\ (\ref{Delta-FF}) with the parameters given there; the 'modified' parameters are given in the text.}
	\label{fig:DeltaFF}
\end{figure}

The discrepancy is most pronounced in the region where the $\Delta$ resonance provides the dominant contribution. We therefore pay now closer attention to the $\Delta$ contribution. Since the cross section is reasonably well described in the very forward bin \cite{Cremonesi:2020} this points to a shortcoming of the $Q^2 = 4  E_\nu E_\mu \sin^2 (\theta/2)$ dependence of the $\Delta$ excitation cross section which should be larger at higher $Q^2$ than given by the original modified dipole form used in GiBUU
\begin{equation}      \label{Delta-FF}
	C_5^\Delta(Q^2) = C_5(0) \left[1 + \frac{aQ^2}{b+Q^2}\right] \left(1 + \frac{Q^2}{{M_A^\Delta}^2}\right)^{-2}
\end{equation}
with $C_5^\Delta = 1.17$, $a = -0.25$, $b = 0.04$ and $M_A^\Delta = 0.95$ GeV \cite{Leitner:2009zz}. In Fig.\ \ref{fig:DeltaFF} we show the $Q^2$-dependence of this original formfactor. In the angular bin just discussed the cross section for the NOvA beam gets its largest contribution for $Q^2 \approx 0.7$ GeV$^2$.

The NOvA collaboration has described their tune of GENIE v 2.12.2 to the data \cite{NOvA:2020rbg}. Since details of the GENIE basis description are not available it is difficult to compare in detail with their results. While the NOvA tune also needs a lowering of the RES contribution at low $Q^2$ the overall $Q^2$ dependence is different from the one found here: we need a lowering of the $\Delta$ formfactor at $Q^2 < 0.05$ GeV$^2$ and an increase at higher $Q^2$. We, therefore  modify the axial $\Delta$ formfactor $C_5^A$ by setting: $a = 0$, $M_A^\Delta = 1.05$ and $C_5^A = 0.85 \cdot 1.17$. This parametrization is close to that used by Hernandez et al. \cite{Hernandez:2007qq}.

For the missing overall strength  we account by increasing the 2p2h and the pion background (BG) contributions by using $\mathcal{T} = 1$ in Eq.\ (\ref{W1nu}).  All these readjustments were done 'by eye'. Remembering that the background contribution in the Bosted-Christy parametrization also contains contributions of broad resonances, in particular for $W > 2$ GeV, the adjustment is within the uncertainties for this region found in Ref.\ \cite{Lalakulich:2006sw}.

\begin{figure}[h]
	\centering
	\includegraphics[width=0.55\linewidth]{"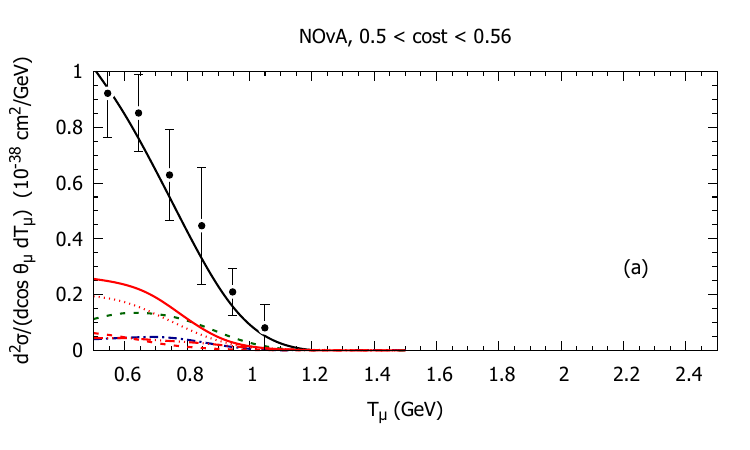"}
	\includegraphics[width=0.55\linewidth]{"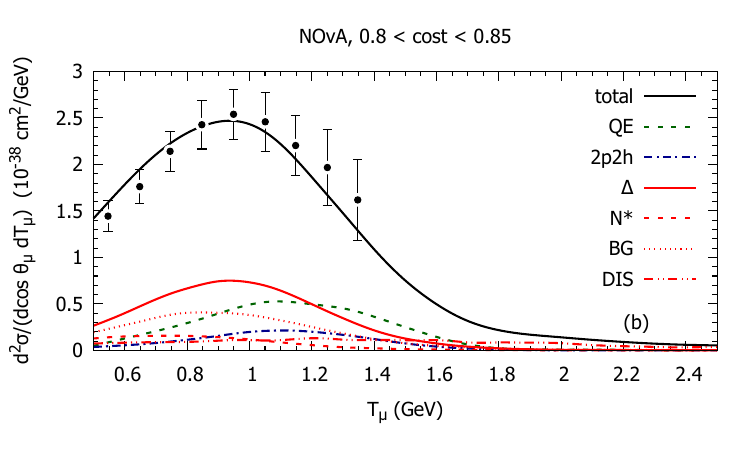"}
	\includegraphics[width=0.55\linewidth]{"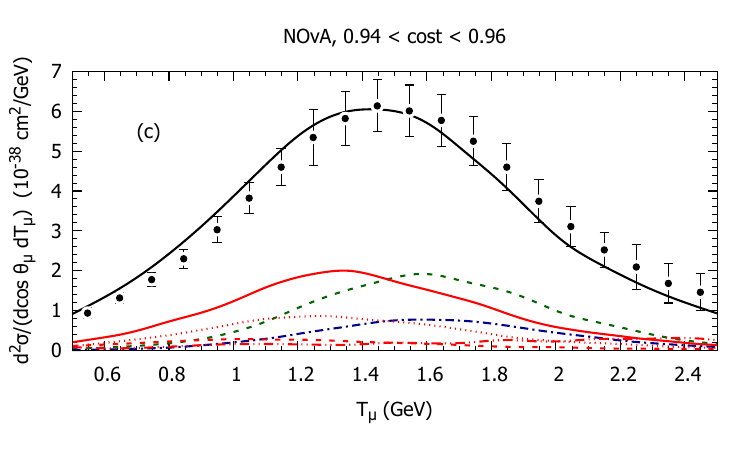"}
		\includegraphics[width=0.55\linewidth]{"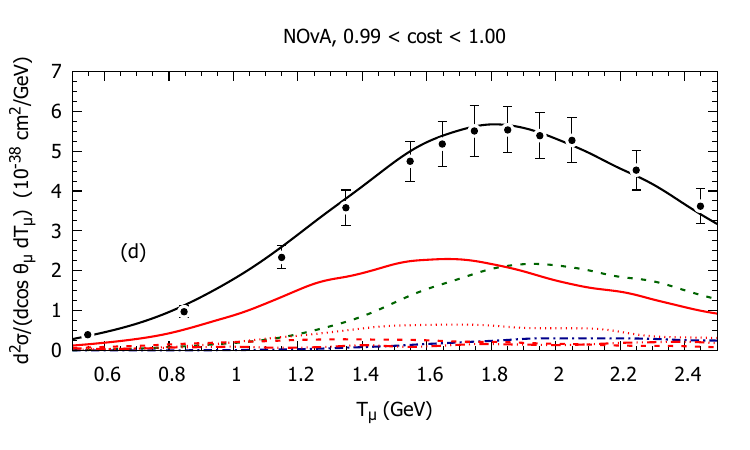"}
	\caption{Double-differential cross sections from the NOvA experiment for various angular bins: (a) $0.5 < \cos \theta < 0.56$, (b)  $0.8 < \cos \theta < 0.85$, (c)  $0.94 < \cos \theta < 0.96$, (d) $0.99 < \cos \theta < 1.00$. Data are taken from \cite{Cremonesi:2020,NOvA:2021eqi}.}
	\label{fig:nova0}
\end{figure}

Fig.\ \ref{fig:nova0} shows the comparison of the results obtained with these modifications for the four angular bins that were shown in \cite{Cremonesi:2020}; there also the comparison with GiBUU results obtained with the version 2019 were shown. The agreement with the data has now significantly been improved.
Noticeable here is that at the very forward direction the cross section is dominated by QE and $\Delta$ excitations, with all other contributions well separated in magnitude. On the other hand, at the most sideways bin (top picture in Fig.\ \ref{fig:nova0}) the background contribution BG reaches the same magnitude as the $\Delta$ contribution reflecting the different $Q^2$-dependencies of these contributions.

In the following sections we now use this tune to the NOvA data ($\Delta$ formfactor hardened, $\mathcal{T} = 1$) without any further change to compare the calculated inclusive cross sections with data from the experiments MicroBooNE and MINERvA, the latter both in the low energy (LE) as well as in its medium energy (ME) run.

\subsection{MicroBooNE inclusive data}

\begin{figure}
	\centering
	\includegraphics[width=0.7\linewidth]{"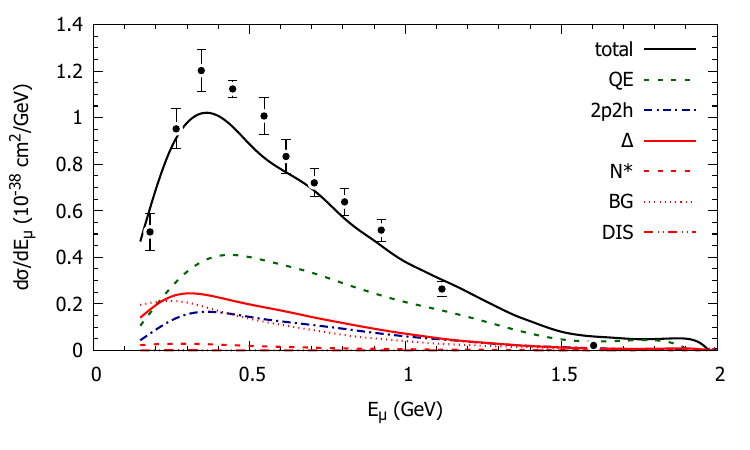"}
	\caption{$E_\mu$ distribution of MicroBooNE. Data are from \cite{MicroBooNE:2021cue}.}
	\label{fig:mcbemu}
\end{figure}

The comparison of the new GiBUU tune with the MicroBooNE data is given in Fig.\ \ref{fig:mcbemu}. In Fig.\ 3 of Ref.\ \cite{MicroBooNE:2021cue} a comparison of the data with various generator predictions is shown also for the $E_\nu$ and the $\omega$ dependencies. In all cases the agreement of the GiBUU results (obtained with v2019) is very good. The present tune used here now does not change that agreement because at the MicroBooNE the largest cross section component is that of QE scattering and RES and DIS contributions play only a minor role. There is still a slight underestimate (about 10\%) in the peak region around $E_\mu \approx 0.35$ GeV, also seen in Fig.\ 3 of Ref.\ \cite{MicroBooNE:2021cue},  and an indication of a somewhat too large contribution at large $E_\mu$ which could be taken as an indication for a slightly larger axial mass for QE scattering and/or an indication for a slightly larger $\Delta$ formfactor at very small $Q^2$.

Further comparisons of GiBUU results with MicroBooNE data can be found in Refs.\ \cite{MicroBooNE:2023cmw,MicroBooNE:2023tzj}.

\subsection{MINERvA LE inclusive data}
The MINERvA low-energy (LE) experiment works with the same neutrino beam as NOvA, but - contrary to NOvA - it sits directly in the beam. The latter leads to a broader energy distribution with a peak at about 3.5 GeV. A comparison of the inclusive data \cite{MINERvA:2020zzv} had shown that v2019 of GiBUU gave a significantly too low cross section. In version 2021 this had been cured by an ad-hoc increase of all cross sections in the SIS region by constant factors.

\begin{figure}[h]
	\centering
	\includegraphics[width=0.7\linewidth]{"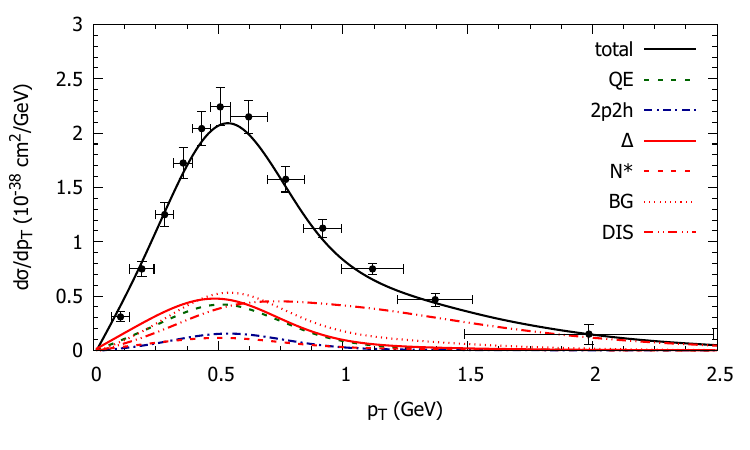"}
	\caption{Transverse muon momentum distribution at the MINERvA LE experiment. The data are taken from \cite{MINERvA:2020zzv}.}
	\label{fig:minervaleldefparms}
\end{figure}

\begin{figure}
	\centering
	\includegraphics[width=0.7\linewidth]{"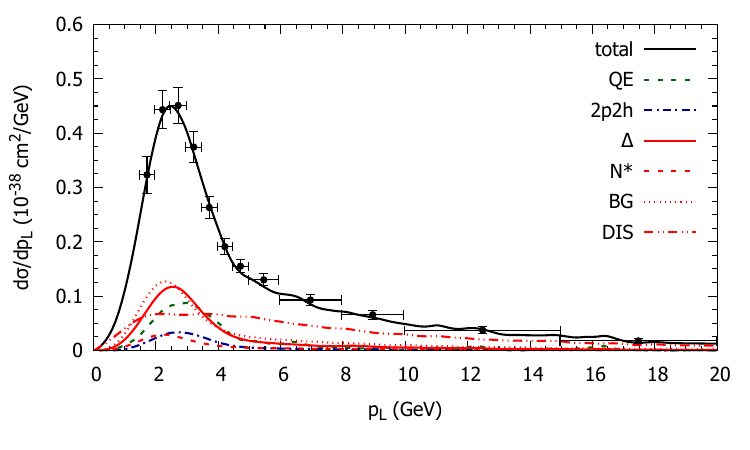"}
	\caption{Longitudinal muon momentum distribution at the MINERvA LE experiment. The data are taken from \cite{MINERvA:2020zzv}.}
	\label{fig:minervaletdefparms}
\end{figure}

In Figs.\ \ref{fig:minervaleldefparms} and \ref{fig:minervaletdefparms} we, therefore, now show the comparison for the longitudinal and transverse momentum distributions obtained with the present version where the $\Delta$ formfactor and the background contribution were chosen such as to optimize the description of the NOvA data. The agreement of the GiBUU results with the data is nearly perfect. 

The DIS contribution extends over the whole momentum range in both figures. For larger $p_T$ and $p_L$ it is the dominant distribution. The analysis of the $W$ distribution in Fig.\ \ref{fig:Wdistr}) shows that these DIS events come mainly from the $W$ region between 2 and 3 GeV, where the BG processes and the true DIS processes overlap. In the peak regions of both distributions the dominant contributions are those of the non-resonant background BG, $\Delta$ excitation and QE scattering. The 2p2h contribution as well as the resonance contribution is small. The BG component is comparable to that for the electron experiment at 2.222 GeV (see Fig.\ \ref{fig:eAr}). For the electron experiment with its fixed energy and angle the invariant mass is limited to the region $W < 1.7$ GeV whereas the inclusive MINERvA LE distribution reaches out to significantly higher $W$ (see Fig.\ \ref{fig:Wdistr}).

\subsection{MINERvA ME inclusive data}
\begin{figure}
	\centering
	\includegraphics[width=0.7\linewidth]{"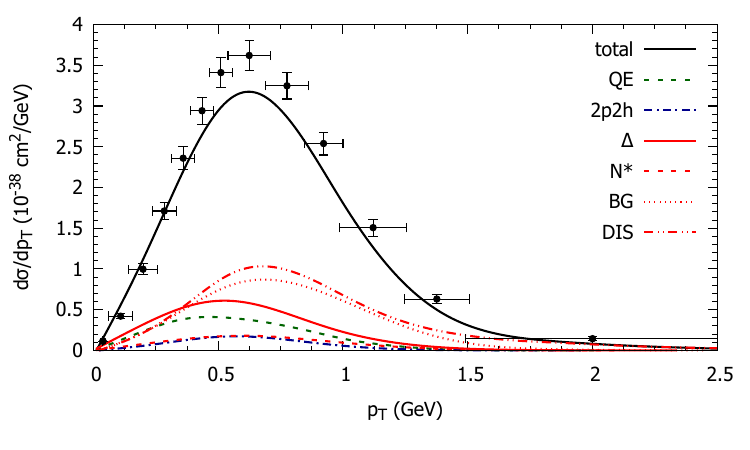"}
	\caption{Transverse muon momentum distribution at the MINERvA ME experiment. The data are taken from \cite{MINERvA:2021owq}.}
	\label{fig:minervamedefl}
\end{figure}

\begin{figure}
	\centering
	\includegraphics[width=0.7\linewidth]{"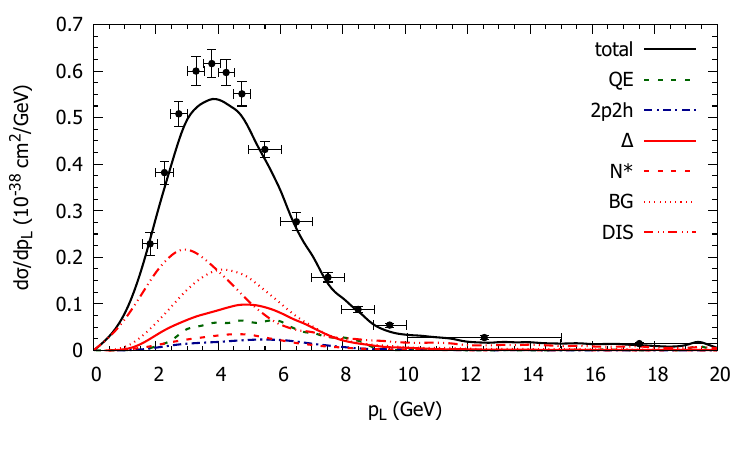"}
	\caption{Longitudinal muon momentum distribution at the MINERvA ME experiment. The data are taken from \cite{MINERvA:2021owq}.}
	\label{fig:minervamedeft}
\end{figure}

Finally we compare the GiBUU results with the MINERvA  medium energy (ME) results, obtained in a higher energy beam \cite{MINERvA:2021owq} with an average energy of 6 GeV. The comparisons in Figs.\ \ref{fig:minervamedefl} and \ref{fig:minervamedeft} show an underestimate (about 10 - 15\%) in the peak regions of both distributions, but otherwise a good agreement. In both distributions now the BG and the DIS contributions are dominant and comparable in magnitude.  This hints again at a strong contribution of excitations in the mass range above about 1.5 GeV, i.e.\ above the $\Delta$ resonance. In th e$p_L$ distribution the DIS peak is shifted to lower $p_L$ with respect to the BG contribution. This reflect the fact that lower $p_L$ are connected with higher target excitations.

The MINERvA collaboration has performed a special tune of GENIE by changing various elementary process contributions. In Figs.\ 9 and 10 in Ref.\ \cite{MINERvA:2021owq} it can be seen that even after that fit ('MINERvA Tune v1') the cross sections are underestimated at the peak by about the same amount as the GiBUU results. The various individual contributions obtained there show some noteable differences to the results here. First, at $p_L \approx 4$ GeV the sum of 'Soft DIS' and 'True DIS' in Fig.\ 9 of \cite{MINERvA:2021owq} is larger than the corresponding sum of BG and DIS obtained here. The comparison of the transverse distributions shows larger differences. Here the MINERvA resonance contribution and the Soft DIS component are larger than the ones obtained here. 

The results of Ref.\ \cite{Gonzalez-Rosa:2023aim} show a much larger underestimate (by about 30\%) at the peak. Closer inspection shows that their TrueDIS + SoftDIS contributions are significantly smaller than the sum of BG and DIS obtained here indicating a problem with the inelastic scaling function and/or the treatment of DIS for $W > 2$ GeV used in the calculations of Ref.\ \cite{Gonzalez-Rosa:2023aim}.  

\subsection{W-distributions}
\begin{figure}
	\centering
	\includegraphics[width=0.7\linewidth]{"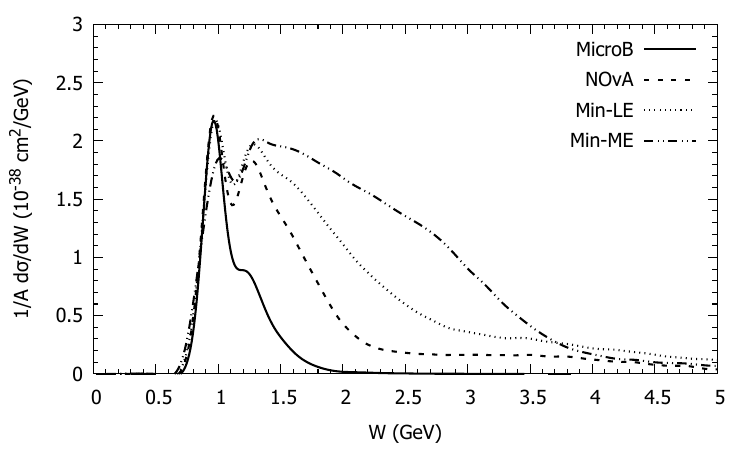"}
	\caption{Distribution of invariant mass $W$ for the MicroBooNE, the NOvA, the MINERvA LE and the MINERvA ME flux for a Carbon target.}
	\label{fig:Wdistr}
\end{figure}

The reason for the underestimate at the peak of the $p$-distributions can be explored by analyzing the $W$ distribution populated in this experiment. Fig.\ \ref{fig:Wdistr} shows that the $\Delta$ resonance and the 2nd resonance region (around the $D_{13}$ N(1520) resonance) are excited in the NOvA, MINERvA LE and MINERvA ME experiments with about the same probability. However, the mass region between about 1.5 and 4.0 GeV is significantly enhanced in the ME experiment. Any missing strength thus presumably comes from that W-region since the MINERvA LE data are described quite well. 
 Problematic is the DIS contribution in this region. DIS in GiBUU is handled by PYTHIA which in itself had been tuned to data at incoming energies of about $> 10$ GeV. We see this as a challenge to retune the PYTHIA parameters to data in the energy regime relevant here.

These results show that in particular the MINERvA ME data contain interesting information on the higher-lying resonances and the so-called SIS/DIS region. Lalakulich and Paschos \cite{Lalakulich:2006sw} have shown that the neutrino cross-sections for resonance excitations can be uncertain by up to a factor of 2. The inclusive data obtained by the MINERvA ME experiment could help to put constraints on these couplings. In the lower energy region ($W < 2$ GeV) the cross section for all experiments (besides MicroBooNE) is sensitive to the nucleon resonance excitations and, even more so, to the background contributions there.  

Furthermore, studies of the Bloom-Gilman effect in nuclei have shown that the integrated strength in the resonance region in nuclei is considerably lower than in the DIS region \cite{Lalakulich:2009zza}. In these calculations only the resonance excitations were considered without any background contribution. The latter would clearly increase the integrated strength and bring the Bloom-Gilman curves into better agreement.

\section{Summary and Conclusions}
In this paper we have summarized the new physics aspects that were recently implemented into the GiBUU theory and generator. The main change lies in the treatment of electron-induced reactions which now starts from the Bosted-Christy parametrizations of the elementary electron-nucleon cross sections. These cross sections are evaluated in the rest frame of the bound Fermi-moving nucleon obtained by a Lorentz-transformation. An important feature of this theory is that as a consequence of the Lorentz transformation the structure functions for the nucleus and the nucleon are evaluated at the same four-momentum transfer $Q^2$, but at different energy- and three-momentum-transfers. The MEC contribution for electrons was obtained from a new fit to electron data by Bodek and Christy \cite{Bodek:2022gli}. 

For neutrinos the QE- and the nucleon-resonance contributions were evaluated by assuming PCAC for the axial current; the vector currents were obtained from the electron fit \cite{Leitner:2006ww,Leitner:2006sp,Leitner:2009zz}. The 2p2h contribution together with the background contribution for neutrinos was obtained by assuming predominance of transverse scattering which allowed to obtain the relevant neutrino structure functions from the transverse structure function for electrons. The theory underlying this transformation from electrons to neutrinos is based on a model of single particle excitations. Since its structure is physically very transparent we have used it also for two-nucleon processes and for possibly more complicated nuclear excitations. The comparison with experimental inclusive data in a wide kinematical regime from MicroBooNE to MINERvA ME shows that indeed it works quite well. 

An essential result of the present study is this success of the transformations from electron to neutrino structure functions given in Eq. (\ref{W1nu}). If this transformation is shown to work also for semi-inclusive data it would provide an enormous simplification. In this case it opens the possibility to consistently transform results obtained in the $e4\nu$ electron-scattering experiments \cite{CLAS:2021neh}, with only an overall normalization, but no tune of the ($Q^2,\omega$) dependence needed.

Within this new implementation we have shown that a very satisfactory description of both electron and neutrino inclusive cross sections can be obtained over a wide range of incoming energies. While the Fermilab Short Baseline and T2K experiments are dominated by QE, 2p2h and $\Delta$ (together with the background) excitations, the MINERvA LE and ME experiments explore the higher energy regions, with NOvA in between. In particular the MINERvA ME experiment is sensitive up to invariant masses of $W \approx 4$ GeV, i.e. well above the resonance region. The cross section there is dominated by the non-resonant background contributions. This observation explains why an earlier study had shown that the Bloom-Gilman duality was not fulfilled for reactions on the nucleus \cite{Lalakulich:2007zz}; only resonance contributions had been considered. It will be interesting to repeat the earlier calculations with the background obtained here taken into account.

Finally, we would like to stress that GiBUU allows the computation of the full final state of a (lepton,nucleus) reaction giving the four-vectors of all final state particles. The final state propagation uses quantum-kinetic transport theory and it employs the same potential $U$ that also appears in the spectral function. There is thus no factorization between initial and final state interactions. The description of all processes (except for the MEC part) is based on a consistent microscopic model that treats initial and final states of the nucleon on the same basis. It does not require further assumptions as in generators which, for example, try to link a model for inclusive cross sections, such as the SUSA approach, to more exclusive final states. The theory implemented in GiBUU also takes into account that the final hadrons start their trajectories somewhere inside the target nucleus because both photons and neutrinos 'illuminate' the whole nucleus (except for shadowing which, however, markedly sets in only at higher energies). This is not the case in the generators using the INTRANUKE package for their treatment of FSI.

\appendix*     
\section{MEC parametrization} \label{MEC}
We give here the expression for the MEC (2p2h) component as used in the present version of the code. The analytic form is taken from \cite{Bodek:2022gli}; this reference contains misprints for the parameters which are corrected here \cite{Christy:2023}.

The transverse structure function $F_1$ is given by
\begin{eqnarray}
	F_1^{\rm MEC} &=& {\rm max}[(f_1^A + f_1^B),0.0] \, \Theta(\omega - \omega_{\rm min}) \nonumber \\
	f_1^A &=& a_1 Y \left[(W^2 - W^2_{\rm min})^{1.5}\, e^{-(W^2 - b_1)^2/(2 c_1^2)}\right] \\
	f_2^A &=& a_2 Y (Q^2 + q_0^2)^{1.5}\, e^{-(W^2 - b_2^2/(2 c_2^2))}
\end{eqnarray}
with
\begin{eqnarray}
	W^2_{\rm min} &=& M^2 + 2M\omega_{\rm min} - Q^2 \nonumber \\
	Y &=& A e^{-Q^4/12.715}\, \frac{\left(Q^2 + q_0^2\right)}{(0.13380 + Q^2)^{6.90679}} \nonumber \\	
	\end{eqnarray}
where $Q^2$ is in units of GeV$^2$, $A$ is the atomic mass number and $M$ the proton mass (0.938 GeV in GiBUU). The parameters are given by $\omega_{\rm min} = 0.0165$ GeV, $q_0^2 = 1.0 \times 10^{-4}, a1 = 0.091648, b_1 = 0.77023, c_1 = 0.077051 + 0.26795\, Q^2, a_2 = 0.01045,  b_2 = 1.275$ and $c_2 = 0.265$.

The MEC structure function $F_1^{\rm MEC}$ is taken here to be proportional to the mass number, reflecting the short-range character of this interaction. For a zero-range force in nuclear matter this linear dependence is exact; a more refined version, which is an option in GiBUU, takes the surface-suppression of light nuclei into account \cite{Mosel:2016uge}.

\begin{acknowledgments} 
We gratefully acknowledge many helpful discussions with Eric Christy about the Bosted-Christy fits to elementary electron-nucleon interactions. We are also grateful to him for providing us some years ago with the code for the Bosted-Christy fits as well as for some discussions of his more recent analysis. We have also benefitted from many discussions with colleagues in the SBND, MINERvA and NOvA experiments.
\end{acknowledgments}

\bibliography{nuclear.bib}

\begin{thebibliography}{76}%
\makeatletter
\providecommand \@ifxundefined [1]{%
 \@ifx{#1\undefined}
}%
\providecommand \@ifnum [1]{%
 \ifnum #1\expandafter \@firstoftwo
 \else \expandafter \@secondoftwo
 \fi
}%
\providecommand \@ifx [1]{%
 \ifx #1\expandafter \@firstoftwo
 \else \expandafter \@secondoftwo
 \fi
}%
\providecommand \natexlab [1]{#1}%
\providecommand \enquote  [1]{``#1''}%
\providecommand \bibnamefont  [1]{#1}%
\providecommand \bibfnamefont [1]{#1}%
\providecommand \citenamefont [1]{#1}%
\providecommand \href@noop [0]{\@secondoftwo}%
\providecommand \href [0]{\begingroup \@sanitize@url \@href}%
\providecommand \@href[1]{\@@startlink{#1}\@@href}%
\providecommand \@@href[1]{\endgroup#1\@@endlink}%
\providecommand \@sanitize@url [0]{\catcode `\\12\catcode `\$12\catcode
  `\&12\catcode `\#12\catcode `\^12\catcode `\_12\catcode `\%12\relax}%
\providecommand \@@startlink[1]{}%
\providecommand \@@endlink[0]{}%
\providecommand \url  [0]{\begingroup\@sanitize@url \@url }%
\providecommand \@url [1]{\endgroup\@href {#1}{\urlprefix }}%
\providecommand \urlprefix  [0]{URL }%
\providecommand \Eprint [0]{\href }%
\providecommand \doibase [0]{https://doi.org/}%
\providecommand \selectlanguage [0]{\@gobble}%
\providecommand \bibinfo  [0]{\@secondoftwo}%
\providecommand \bibfield  [0]{\@secondoftwo}%
\providecommand \translation [1]{[#1]}%
\providecommand \BibitemOpen [0]{}%
\providecommand \bibitemStop [0]{}%
\providecommand \bibitemNoStop [0]{.\EOS\space}%
\providecommand \EOS [0]{\spacefactor3000\relax}%
\providecommand \BibitemShut  [1]{\csname bibitem#1\endcsname}%
\let\auto@bib@innerbib\@empty
\bibitem [{\citenamefont {Drechsel}\ and\ \citenamefont
  {Giannini}(1989)}]{Drechsel:1989ab}%
  \BibitemOpen
  \bibfield  {author} {\bibinfo {author} {\bibfnamefont {D.}~\bibnamefont
  {Drechsel}}\ and\ \bibinfo {author} {\bibfnamefont {M.~M.}\ \bibnamefont
  {Giannini}},\ }\bibfield  {title} {\bibinfo {title} {{ELECTRON SCATTERING OFF
  NUCLEI}},\ }\href {https://doi.org/10.1088/0034-4885/52/9/002} {\bibfield
  {journal} {\bibinfo  {journal} {Rept. Prog. Phys.}\ }\textbf {\bibinfo
  {volume} {52}},\ \bibinfo {pages} {1083} (\bibinfo {year}
  {1989})}\BibitemShut {NoStop}%
\bibitem [{\citenamefont {Boffi}\ \emph {et~al.}(1996)\citenamefont {Boffi},
  \citenamefont {Giusti}, \citenamefont {Pacati},\ and\ \citenamefont
  {Radici}}]{Boffi:1996ikg}%
  \BibitemOpen
  \bibfield  {author} {\bibinfo {author} {\bibfnamefont {S.}~\bibnamefont
  {Boffi}}, \bibinfo {author} {\bibfnamefont {C.}~\bibnamefont {Giusti}},
  \bibinfo {author} {\bibfnamefont {F.~d.}\ \bibnamefont {Pacati}},\ and\
  \bibinfo {author} {\bibfnamefont {M.}~\bibnamefont {Radici}},\ }\href@noop {}
  {\emph {\bibinfo {title} {{Electromagnetic Response of Atomic Nuclei}}}},\
  \bibinfo {series} {Oxford Studies in Nuclear Physics}, Vol.~\bibinfo {volume}
  {20}\ (\bibinfo  {publisher} {Clarendon Press},\ \bibinfo {address} {Oxford
  UK},\ \bibinfo {year} {1996})\BibitemShut {NoStop}%
\bibitem [{\citenamefont {Gallagher}\ \emph {et~al.}(2011)\citenamefont
  {Gallagher}, \citenamefont {Garvey},\ and\ \citenamefont
  {Zeller}}]{Gallagher:2011zza}%
  \BibitemOpen
  \bibfield  {author} {\bibinfo {author} {\bibfnamefont {H.}~\bibnamefont
  {Gallagher}}, \bibinfo {author} {\bibfnamefont {G.}~\bibnamefont {Garvey}},\
  and\ \bibinfo {author} {\bibfnamefont {G.}~\bibnamefont {Zeller}},\
  }\bibfield  {title} {\bibinfo {title} {{Neutrino-nucleus interactions}},\
  }\href {https://doi.org/10.1146/annurev-nucl-102010-130255} {\bibfield
  {journal} {\bibinfo  {journal} {Ann.Rev.Nucl.Part.Sci.}\ }\textbf {\bibinfo
  {volume} {61}},\ \bibinfo {pages} {355} (\bibinfo {year} {2011})}\BibitemShut
  {NoStop}%
\bibitem [{\citenamefont {Diwan}\ \emph {et~al.}(2016)\citenamefont {Diwan},
  \citenamefont {Galymov}, \citenamefont {Qian},\ and\ \citenamefont
  {Rubbia}}]{Diwan:2016gmz}%
  \BibitemOpen
  \bibfield  {author} {\bibinfo {author} {\bibfnamefont {M.~V.}\ \bibnamefont
  {Diwan}}, \bibinfo {author} {\bibfnamefont {V.}~\bibnamefont {Galymov}},
  \bibinfo {author} {\bibfnamefont {X.}~\bibnamefont {Qian}},\ and\ \bibinfo
  {author} {\bibfnamefont {A.}~\bibnamefont {Rubbia}},\ }\bibfield  {title}
  {\bibinfo {title} {{Long-Baseline Neutrino Experiments}},\ }\href
  {https://doi.org/10.1146/annurev-nucl-102014-021939} {\bibfield  {journal}
  {\bibinfo  {journal} {Ann. Rev. Nucl. Part. Sci.}\ }\textbf {\bibinfo
  {volume} {66}},\ \bibinfo {pages} {47} (\bibinfo {year} {2016})},\ \Eprint
  {https://arxiv.org/abs/1608.06237} {arXiv:1608.06237 [hep-ex]} \BibitemShut
  {NoStop}%
\bibitem [{\citenamefont {Katori}\ and\ \citenamefont
  {Martini}(2018)}]{Katori:2016yel}%
  \BibitemOpen
  \bibfield  {author} {\bibinfo {author} {\bibfnamefont {T.}~\bibnamefont
  {Katori}}\ and\ \bibinfo {author} {\bibfnamefont {M.}~\bibnamefont
  {Martini}},\ }\bibfield  {title} {\bibinfo {title} {{Neutrino-nucleus cross
  sections for oscillation experiments}},\ }\href
  {https://doi.org/10.1088/1361-6471/aa8bf7} {\bibfield  {journal} {\bibinfo
  {journal} {J. Phys.}\ }\textbf {\bibinfo {volume} {G45}},\ \bibinfo {pages}
  {013001} (\bibinfo {year} {2018})},\ \Eprint
  {https://arxiv.org/abs/1611.07770} {arXiv:1611.07770 [hep-ph]} \BibitemShut
  {NoStop}%
\bibitem [{\citenamefont {Mosel}(2016)}]{Mosel:2016cwa}%
  \BibitemOpen
  \bibfield  {author} {\bibinfo {author} {\bibfnamefont {U.}~\bibnamefont
  {Mosel}},\ }\bibfield  {title} {\bibinfo {title} {{Neutrino Interactions with
  Nucleons and Nuclei: Importance for Long-Baseline Experiments}},\ }\href
  {https://doi.org/10.1146/annurev-nucl-102115-044720} {\bibfield  {journal}
  {\bibinfo  {journal} {Ann. Rev. Nucl. Part. Sci.}\ }\textbf {\bibinfo
  {volume} {66}},\ \bibinfo {pages} {171} (\bibinfo {year} {2016})},\ \Eprint
  {https://arxiv.org/abs/1602.00696} {arXiv:1602.00696 [nucl-th]} \BibitemShut
  {NoStop}%
\bibitem [{\citenamefont {Mosel}(2019)}]{Mosel:2019vhx}%
  \BibitemOpen
  \bibfield  {author} {\bibinfo {author} {\bibfnamefont {U.}~\bibnamefont
  {Mosel}},\ }\bibfield  {title} {\bibinfo {title} {{Neutrino event generators:
  foundation, status and future}},\ }\href
  {https://doi.org/10.1088/1361-6471/ab3830} {\bibfield  {journal} {\bibinfo
  {journal} {J. Phys. G}\ }\textbf {\bibinfo {volume} {46}},\ \bibinfo {pages}
  {113001} (\bibinfo {year} {2019})},\ \Eprint
  {https://arxiv.org/abs/1904.11506} {arXiv:1904.11506 [hep-ex]} \BibitemShut
  {NoStop}%
\bibitem [{\citenamefont {Gallagher}\ and\ \citenamefont
  {Hayato}()}]{NeutrinoGen:2022}%
  \BibitemOpen
  \bibfield  {author} {\bibinfo {author} {\bibfnamefont {H.}~\bibnamefont
  {Gallagher}}\ and\ \bibinfo {author} {\bibfnamefont {Y.}~\bibnamefont
  {Hayato}},\ }\bibinfo {title} {Monte carlo neutrino generators},\ in\
  \href@noop {} {\emph {\bibinfo {booktitle} {Particle Data Group}}},\ \bibinfo
  {editor} {edited by\ \bibinfo {editor} {\bibfnamefont {R.}~\bibnamefont
  {Workman}}\ and\ \bibinfo {editor} {\bibnamefont {et~al.}}}\ (\bibinfo
  {publisher} {Prog. Theor. Exp. Phys. 2022, 083C01})\BibitemShut {NoStop}%
\bibitem [{\citenamefont {Bronner}\ and\ \citenamefont
  {Hartz}(2016)}]{Bronner:2016gmz}%
  \BibitemOpen
  \bibfield  {author} {\bibinfo {author} {\bibfnamefont {C.}~\bibnamefont
  {Bronner}}\ and\ \bibinfo {author} {\bibfnamefont {M.}~\bibnamefont
  {Hartz}},\ }\bibfield  {title} {\bibinfo {title} {{Tuning of the Charged
  Hadrons Multiplicities for Deep Inelastic Interactions in NEUT}},\ }\href
  {https://doi.org/10.7566/JPSCP.12.010041} {\bibfield  {journal} {\bibinfo
  {journal} {JPS Conf. Proc.}\ }\textbf {\bibinfo {volume} {12}},\ \bibinfo
  {pages} {010041} (\bibinfo {year} {2016})},\ \Eprint
  {https://arxiv.org/abs/1607.06558} {arXiv:1607.06558 [hep-ph]} \BibitemShut
  {NoStop}%
\bibitem [{\citenamefont {Tena-Vidal}\ \emph
  {et~al.}(2021{\natexlab{a}})\citenamefont {Tena-Vidal} \emph
  {et~al.}}]{Tena-Vidal:2021rpu}%
  \BibitemOpen
  \bibfield  {author} {\bibinfo {author} {\bibfnamefont {J.}~\bibnamefont
  {Tena-Vidal}} \emph {et~al.} (\bibinfo {collaboration} {GENIE}),\ }\bibfield
  {title} {\bibinfo {title} {{Neutrino-Nucleon Cross-Section Model Tuning in
  GENIE v3}},\ }\href@noop {} {\  (\bibinfo {year} {2021}{\natexlab{a}})},\
  \Eprint {https://arxiv.org/abs/2104.09179} {arXiv:2104.09179 [hep-ph]}
  \BibitemShut {NoStop}%
\bibitem [{\citenamefont {Tena-Vidal}\ \emph
  {et~al.}(2021{\natexlab{b}})\citenamefont {Tena-Vidal} \emph
  {et~al.}}]{GENIE:2021wox}%
  \BibitemOpen
  \bibfield  {author} {\bibinfo {author} {\bibfnamefont {J.}~\bibnamefont
  {Tena-Vidal}} \emph {et~al.} (\bibinfo {collaboration} {GENIE}),\ }\bibfield
  {title} {\bibinfo {title} {{Hadronization Model Tuning in GENIE v3}},\
  }\href@noop {} {\  (\bibinfo {year} {2021}{\natexlab{b}})},\ \Eprint
  {https://arxiv.org/abs/2106.05884} {arXiv:2106.05884 [hep-ph]} \BibitemShut
  {NoStop}%
\bibitem [{\citenamefont {Abratenko}\ \emph
  {et~al.}(2021{\natexlab{a}})\citenamefont {Abratenko} \emph
  {et~al.}}]{MicroBooNE:2021ccs}%
  \BibitemOpen
  \bibfield  {author} {\bibinfo {author} {\bibfnamefont {P.}~\bibnamefont
  {Abratenko}} \emph {et~al.} (\bibinfo {collaboration} {MicroBooNE}),\
  }\bibfield  {title} {\bibinfo {title} {{New Theory-driven GENIE Tune for
  MicroBooNE}},\ }\href@noop {} {\  (\bibinfo {year} {2021}{\natexlab{a}})},\
  \Eprint {https://arxiv.org/abs/2110.14028} {arXiv:2110.14028 [hep-ex]}
  \BibitemShut {NoStop}%
\bibitem [{\citenamefont {Acero}\ \emph {et~al.}(2020)\citenamefont {Acero}
  \emph {et~al.}}]{NOvA:2020rbg}%
  \BibitemOpen
  \bibfield  {author} {\bibinfo {author} {\bibfnamefont {M.~A.}\ \bibnamefont
  {Acero}} \emph {et~al.} (\bibinfo {collaboration} {NOvA, R. Group}),\
  }\bibfield  {title} {\bibinfo {title} {{Adjusting neutrino interaction models
  and evaluating uncertainties using NOvA near detector data}},\ }\href
  {https://doi.org/10.1140/epjc/s10052-020-08577-5} {\bibfield  {journal}
  {\bibinfo  {journal} {Eur. Phys. J. C}\ }\textbf {\bibinfo {volume} {80}},\
  \bibinfo {pages} {1119} (\bibinfo {year} {2020})},\ \Eprint
  {https://arxiv.org/abs/2006.08727} {arXiv:2006.08727 [hep-ex]} \BibitemShut
  {NoStop}%
\bibitem [{\citenamefont {Aguilar-Arevalo}\ \emph {et~al.}(2008)\citenamefont
  {Aguilar-Arevalo} \emph {et~al.}}]{MiniBooNE:2007iti}%
  \BibitemOpen
  \bibfield  {author} {\bibinfo {author} {\bibfnamefont {A.~A.}\ \bibnamefont
  {Aguilar-Arevalo}} \emph {et~al.} (\bibinfo {collaboration} {MiniBooNE}),\
  }\bibfield  {title} {\bibinfo {title} {{Measurement of muon neutrino
  quasi-elastic scattering on carbon}},\ }\href
  {https://doi.org/10.1103/PhysRevLett.100.032301} {\bibfield  {journal}
  {\bibinfo  {journal} {Phys. Rev. Lett.}\ }\textbf {\bibinfo {volume} {100}},\
  \bibinfo {pages} {032301} (\bibinfo {year} {2008})},\ \Eprint
  {https://arxiv.org/abs/0706.0926} {arXiv:0706.0926 [hep-ex]} \BibitemShut
  {NoStop}%
\bibitem [{\citenamefont {Martini}\ \emph {et~al.}(2009)\citenamefont
  {Martini}, \citenamefont {Ericson}, \citenamefont {Chanfray},\ and\
  \citenamefont {Marteau}}]{Martini:2009uj}%
  \BibitemOpen
  \bibfield  {author} {\bibinfo {author} {\bibfnamefont {M.}~\bibnamefont
  {Martini}}, \bibinfo {author} {\bibfnamefont {M.}~\bibnamefont {Ericson}},
  \bibinfo {author} {\bibfnamefont {G.}~\bibnamefont {Chanfray}},\ and\
  \bibinfo {author} {\bibfnamefont {J.}~\bibnamefont {Marteau}},\ }\bibfield
  {title} {\bibinfo {title} {{A Unified approach for nucleon knock-out,
  coherent and incoherent pion production in neutrino interactions with
  nuclei}},\ }\href {https://doi.org/10.1103/PhysRevC.80.065501} {\bibfield
  {journal} {\bibinfo  {journal} {Phys.Rev.C}\ }\textbf {\bibinfo {volume}
  {80}},\ \bibinfo {pages} {065501} (\bibinfo {year} {2009})},\ \Eprint
  {https://arxiv.org/abs/0910.2622} {arXiv:0910.2622 [nucl-th]} \BibitemShut
  {NoStop}%
\bibitem [{\citenamefont {Nieves}\ \emph {et~al.}(2011)\citenamefont {Nieves},
  \citenamefont {Ruiz~Simo},\ and\ \citenamefont
  {Vicente~Vacas}}]{Nieves:2011pp}%
  \BibitemOpen
  \bibfield  {author} {\bibinfo {author} {\bibfnamefont {J.}~\bibnamefont
  {Nieves}}, \bibinfo {author} {\bibfnamefont {I.}~\bibnamefont {Ruiz~Simo}},\
  and\ \bibinfo {author} {\bibfnamefont {M.}~\bibnamefont {Vicente~Vacas}},\
  }\bibfield  {title} {\bibinfo {title} {{Inclusive Charged--Current
  Neutrino--Nucleus Reactions}},\ }\href
  {https://doi.org/10.1103/PhysRevC.83.045501} {\bibfield  {journal} {\bibinfo
  {journal} {Phys.Rev.C}\ }\textbf {\bibinfo {volume} {83}},\ \bibinfo {pages}
  {045501} (\bibinfo {year} {2011})},\ \Eprint
  {https://arxiv.org/abs/1102.2777} {arXiv:1102.2777 [hep-ph]} \BibitemShut
  {NoStop}%
\bibitem [{{GiBUU Website}()}]{gibuu}%
  \BibitemOpen
  {GiBUU Website},\ \href@noop {} {}\bibinfo {note}
  {\url{http://gibuu.hepforge.org}}\BibitemShut {NoStop}%
\bibitem [{\citenamefont {Leitner}(2009)}]{Leitner:2009zz}%
  \BibitemOpen
  \bibfield  {author} {\bibinfo {author} {\bibfnamefont {T.~J.}\ \bibnamefont
  {Leitner}},\ }\emph {\bibinfo {title} {{Neutrino-nucleus interactions in a
  coupled-channel hadronic transport model}}},\ \href@noop {} {\bibinfo {type}
  {Phd thesis}},\ \bibinfo  {school} {{JLU Giessen}}, \bibinfo {address}
  {{Giessen, Germany}} (\bibinfo {year} {2009}),\ \bibinfo {note}
  {https://inspirehep.net/files/af05a724f4c31c0070e8033fd1140ad2}\BibitemShut
  {NoStop}%
\bibitem [{\citenamefont {Buss}\ \emph {et~al.}(2012)\citenamefont {Buss},
  \citenamefont {Gaitanos}, \citenamefont {Gallmeister}, \citenamefont {van
  Hees}, \citenamefont {Kaskulov} \emph {et~al.}}]{Buss:2011mx}%
  \BibitemOpen
  \bibfield  {author} {\bibinfo {author} {\bibfnamefont {O.}~\bibnamefont
  {Buss}}, \bibinfo {author} {\bibfnamefont {T.}~\bibnamefont {Gaitanos}},
  \bibinfo {author} {\bibfnamefont {K.}~\bibnamefont {Gallmeister}}, \bibinfo
  {author} {\bibfnamefont {H.}~\bibnamefont {van Hees}}, \bibinfo {author}
  {\bibfnamefont {M.}~\bibnamefont {Kaskulov}}, \emph {et~al.},\ }\bibfield
  {title} {\bibinfo {title} {{Transport-theoretical Description of Nuclear
  Reactions}},\ }\href {https://doi.org/10.1016/j.physrep.2011.12.001}
  {\bibfield  {journal} {\bibinfo  {journal} {Phys.Rept.}\ }\textbf {\bibinfo
  {volume} {512}},\ \bibinfo {pages} {1} (\bibinfo {year} {2012})},\ \Eprint
  {https://arxiv.org/abs/1106.1344} {arXiv:1106.1344 [hep-ph]} \BibitemShut
  {NoStop}%
\bibitem [{\citenamefont {Gallmeister}\ \emph {et~al.}(2016)\citenamefont
  {Gallmeister}, \citenamefont {Mosel},\ and\ \citenamefont
  {Weil}}]{Gallmeister:2016dnq}%
  \BibitemOpen
  \bibfield  {author} {\bibinfo {author} {\bibfnamefont {K.}~\bibnamefont
  {Gallmeister}}, \bibinfo {author} {\bibfnamefont {U.}~\bibnamefont {Mosel}},\
  and\ \bibinfo {author} {\bibfnamefont {J.}~\bibnamefont {Weil}},\ }\bibfield
  {title} {\bibinfo {title} {{Neutrino-Induced Reactions on Nuclei}},\ }\href
  {https://doi.org/10.1103/PhysRevC.94.035502} {\bibfield  {journal} {\bibinfo
  {journal} {Phys. Rev.}\ }\textbf {\bibinfo {volume} {C94}},\ \bibinfo {pages}
  {035502} (\bibinfo {year} {2016})},\ \Eprint
  {https://arxiv.org/abs/1605.09391} {arXiv:1605.09391 [nucl-th]} \BibitemShut
  {NoStop}%
\bibitem [{\citenamefont {Gonzalez-Rosa}\ \emph {et~al.}(2023)\citenamefont
  {Gonzalez-Rosa}, \citenamefont {Megias}, \citenamefont {Caballero},\ and\
  \citenamefont {Barbaro}}]{Gonzalez-Rosa:2023aim}%
  \BibitemOpen
  \bibfield  {author} {\bibinfo {author} {\bibfnamefont {J.}~\bibnamefont
  {Gonzalez-Rosa}}, \bibinfo {author} {\bibfnamefont {G.~D.}\ \bibnamefont
  {Megias}}, \bibinfo {author} {\bibfnamefont {J.~A.}\ \bibnamefont
  {Caballero}},\ and\ \bibinfo {author} {\bibfnamefont {M.~B.}\ \bibnamefont
  {Barbaro}},\ }\bibfield  {title} {\bibinfo {title} {{Superscaling in the
  resonance region for neutrino-nucleus scattering: The SuSAv2-DCC model}},\
  }\href@noop {} {\  (\bibinfo {year} {2023})},\ \Eprint
  {https://arxiv.org/abs/2306.12060} {arXiv:2306.12060 [nucl-th]} \BibitemShut
  {NoStop}%
\bibitem [{\citenamefont {Filkins}\ \emph {et~al.}(2020)\citenamefont {Filkins}
  \emph {et~al.}}]{MINERvA:2020zzv}%
  \BibitemOpen
  \bibfield  {author} {\bibinfo {author} {\bibfnamefont {A.}~\bibnamefont
  {Filkins}} \emph {et~al.} (\bibinfo {collaboration} {MINERvA}),\ }\bibfield
  {title} {\bibinfo {title} {{Double-differential inclusive charged-current
  $\nu_\mu$ cross sections on hydrocarbon in MINERvA at $\langle E_{\nu}
  \rangle \sim$ 3.5 GeV}},\ }\href
  {https://doi.org/10.1103/PhysRevD.101.112007} {\bibfield  {journal} {\bibinfo
   {journal} {Phys. Rev. D}\ }\textbf {\bibinfo {volume} {101}},\ \bibinfo
  {pages} {112007} (\bibinfo {year} {2020})},\ \Eprint
  {https://arxiv.org/abs/2002.12496} {arXiv:2002.12496 [hep-ex]} \BibitemShut
  {NoStop}%
\bibitem [{\citenamefont {Ruterbories}\ \emph {et~al.}(2021)\citenamefont
  {Ruterbories} \emph {et~al.}}]{MINERvA:2021owq}%
  \BibitemOpen
  \bibfield  {author} {\bibinfo {author} {\bibfnamefont {D.}~\bibnamefont
  {Ruterbories}} \emph {et~al.} (\bibinfo {collaboration} {MINERvA}),\
  }\bibfield  {title} {\bibinfo {title} {{Measurement of inclusive
  charged-current $\nu_\mu$ cross sections as a function of muon kinematics at
  $<E_\nu>\sim6~GeV$ on hydrocarbon}},\ }\href
  {https://doi.org/10.1103/PhysRevD.104.092007} {\bibfield  {journal} {\bibinfo
   {journal} {Phys. Rev. D}\ }\textbf {\bibinfo {volume} {104}},\ \bibinfo
  {pages} {092007} (\bibinfo {year} {2021})},\ \Eprint
  {https://arxiv.org/abs/2106.16210} {arXiv:2106.16210 [hep-ex]} \BibitemShut
  {NoStop}%
\bibitem [{\citenamefont {Gallmeister}\ and\ \citenamefont
  {Mosel}(2009)}]{Gallmeister:2009ht}%
  \BibitemOpen
  \bibfield  {author} {\bibinfo {author} {\bibfnamefont {K.}~\bibnamefont
  {Gallmeister}}\ and\ \bibinfo {author} {\bibfnamefont {U.}~\bibnamefont
  {Mosel}},\ }\bibfield  {title} {\bibinfo {title} {{Production of charged
  pions off nuclei with 3...30 GeV incident protons and pions}},\ }\href
  {https://doi.org/10.1016/j.nuclphysa.2009.05.104} {\bibfield  {journal}
  {\bibinfo  {journal} {Nucl. Phys.}\ }\textbf {\bibinfo {volume} {A826}},\
  \bibinfo {pages} {151} (\bibinfo {year} {2009})},\ \Eprint
  {https://arxiv.org/abs/0901.1770} {arXiv:0901.1770 [hep-ex]} \BibitemShut
  {NoStop}%
\bibitem [{\citenamefont {Larionov}\ \emph {et~al.}(2021)\citenamefont
  {Larionov}, \citenamefont {Mosel},\ and\ \citenamefont {von
  Smekal}}]{Larionov:2020fnu}%
  \BibitemOpen
  \bibfield  {author} {\bibinfo {author} {\bibfnamefont {A.~B.}\ \bibnamefont
  {Larionov}}, \bibinfo {author} {\bibfnamefont {U.}~\bibnamefont {Mosel}},\
  and\ \bibinfo {author} {\bibfnamefont {L.}~\bibnamefont {von Smekal}},\
  }\bibfield  {title} {\bibinfo {title} {{Dilepton production in microscopic
  transport theory with in-medium $\rho$-meson spectral function}},\ }\href
  {https://doi.org/10.1103/PhysRevC.102.064913} {\bibfield  {journal} {\bibinfo
   {journal} {Phys. Rev. C}\ }\textbf {\bibinfo {volume} {102}},\ \bibinfo
  {pages} {064913} (\bibinfo {year} {2021})},\ \Eprint
  {https://arxiv.org/abs/2009.11702} {arXiv:2009.11702 [nucl-th]} \BibitemShut
  {NoStop}%
\bibitem [{\citenamefont {Lehr}\ and\ \citenamefont
  {Mosel}(2004)}]{Lehr:2003ht}%
  \BibitemOpen
  \bibfield  {author} {\bibinfo {author} {\bibfnamefont {J.}~\bibnamefont
  {Lehr}}\ and\ \bibinfo {author} {\bibfnamefont {U.}~\bibnamefont {Mosel}},\
  }\bibfield  {title} {\bibinfo {title} {{Influence of the nucleon spectral
  function in photon and electron induced reactions on nuclei}},\ }\href
  {https://doi.org/10.1103/PhysRevC.69.024603} {\bibfield  {journal} {\bibinfo
  {journal} {Phys. Rev. C}\ }\textbf {\bibinfo {volume} {69}},\ \bibinfo
  {pages} {024603} (\bibinfo {year} {2004})},\ \Eprint
  {https://arxiv.org/abs/nucl-th/0309084} {arXiv:nucl-th/0309084} \BibitemShut
  {NoStop}%
\bibitem [{\citenamefont {Buss}\ \emph {et~al.}(2006)\citenamefont {Buss},
  \citenamefont {Alvarez-Ruso}, \citenamefont {Muhlich},\ and\ \citenamefont
  {Mosel}}]{Buss:2006vh}%
  \BibitemOpen
  \bibfield  {author} {\bibinfo {author} {\bibfnamefont {O.}~\bibnamefont
  {Buss}}, \bibinfo {author} {\bibfnamefont {L.}~\bibnamefont {Alvarez-Ruso}},
  \bibinfo {author} {\bibfnamefont {P.}~\bibnamefont {Muhlich}},\ and\ \bibinfo
  {author} {\bibfnamefont {U.}~\bibnamefont {Mosel}},\ }\bibfield  {title}
  {\bibinfo {title} {{Low-energy pions in nuclear matter and pi pi
  photoproduction within a BUU transport model}},\ }\href
  {https://doi.org/10.1140/epja/i2006-10075-y} {\bibfield  {journal} {\bibinfo
  {journal} {Eur.Phys.J.}\ }\textbf {\bibinfo {volume} {A29}},\ \bibinfo
  {pages} {189} (\bibinfo {year} {2006})},\ \Eprint
  {https://arxiv.org/abs/nucl-th/0603003} {arXiv:nucl-th/0603003 [nucl-th]}
  \BibitemShut {NoStop}%
\bibitem [{\citenamefont {Kadanoff}\ and\ \citenamefont
  {Baym}(1962)}]{Kad-Baym:1962}%
  \BibitemOpen
  \bibfield  {author} {\bibinfo {author} {\bibfnamefont {L.}~\bibnamefont
  {Kadanoff}}\ and\ \bibinfo {author} {\bibfnamefont {G.}~\bibnamefont
  {Baym}},\ }\href@noop {} {\emph {\bibinfo {title} {Quantum statistical
  mechanics}}}\ (\bibinfo  {publisher} {Benjamin},\ \bibinfo {address} {New
  York},\ \bibinfo {year} {1962})\BibitemShut {NoStop}%
\bibitem [{\citenamefont {Pandey}\ \emph {et~al.}(2016)\citenamefont {Pandey},
  \citenamefont {Jachowicz}, \citenamefont {Martini}, \citenamefont
  {Gonzalez-Jimenez}, \citenamefont {Ryckebusch}, \citenamefont {Van~Cuyck},\
  and\ \citenamefont {Van~Dessel}}]{Pandey:2016jju}%
  \BibitemOpen
  \bibfield  {author} {\bibinfo {author} {\bibfnamefont {V.}~\bibnamefont
  {Pandey}}, \bibinfo {author} {\bibfnamefont {N.}~\bibnamefont {Jachowicz}},
  \bibinfo {author} {\bibfnamefont {M.}~\bibnamefont {Martini}}, \bibinfo
  {author} {\bibfnamefont {R.}~\bibnamefont {Gonzalez-Jimenez}}, \bibinfo
  {author} {\bibfnamefont {J.}~\bibnamefont {Ryckebusch}}, \bibinfo {author}
  {\bibfnamefont {T.}~\bibnamefont {Van~Cuyck}},\ and\ \bibinfo {author}
  {\bibfnamefont {N.}~\bibnamefont {Van~Dessel}},\ }\bibfield  {title}
  {\bibinfo {title} {{Impact of low-energy nuclear excitations on
  neutrino-nucleus scattering at MiniBooNE and T2K kinematics}},\ }\href
  {https://doi.org/10.1103/PhysRevC.94.054609} {\bibfield  {journal} {\bibinfo
  {journal} {Phys. Rev.}\ }\textbf {\bibinfo {volume} {C94}},\ \bibinfo {pages}
  {054609} (\bibinfo {year} {2016})},\ \Eprint
  {https://arxiv.org/abs/1607.01216} {arXiv:1607.01216 [nucl-th]} \BibitemShut
  {NoStop}%
\bibitem [{\citenamefont {Nieves}\ and\ \citenamefont
  {Sobczyk}(2017)}]{Nieves:2017lij}%
  \BibitemOpen
  \bibfield  {author} {\bibinfo {author} {\bibfnamefont {J.}~\bibnamefont
  {Nieves}}\ and\ \bibinfo {author} {\bibfnamefont {J.~E.}\ \bibnamefont
  {Sobczyk}},\ }\bibfield  {title} {\bibinfo {title} {{In medium dispersion
  relation effects in nuclear inclusive reactions at intermediate and low
  energies}},\ }\href {https://doi.org/10.1016/j.aop.2017.06.002} {\bibfield
  {journal} {\bibinfo  {journal} {Annals Phys.}\ }\textbf {\bibinfo {volume}
  {383}},\ \bibinfo {pages} {455} (\bibinfo {year} {2017})},\ \Eprint
  {https://arxiv.org/abs/1701.03628} {arXiv:1701.03628 [nucl-th]} \BibitemShut
  {NoStop}%
\bibitem [{\citenamefont {Christy}(2015)}]{Christy:2015}%
  \BibitemOpen
  \bibfield  {author} {\bibinfo {author} {\bibfnamefont {E.}~\bibnamefont
  {Christy}},\ }\href@noop {} {}\bibinfo {howpublished} {private communication}
  (\bibinfo {year} {2015})\BibitemShut {NoStop}%
\bibitem [{\citenamefont {Bodek}\ and\ \citenamefont
  {Christy}(2022)}]{Bodek:2022gli}%
  \BibitemOpen
  \bibfield  {author} {\bibinfo {author} {\bibfnamefont {A.}~\bibnamefont
  {Bodek}}\ and\ \bibinfo {author} {\bibfnamefont {M.~E.}\ \bibnamefont
  {Christy}},\ }\bibfield  {title} {\bibinfo {title} {{Extraction of the
  Coulomb sum rule, transverse enhancement, and longitudinal quenching from an
  analysis of all available e-C12 and e-O16 cross section~data}},\ }\href
  {https://doi.org/10.1103/PhysRevC.106.L061305} {\bibfield  {journal}
  {\bibinfo  {journal} {Phys. Rev. C}\ }\textbf {\bibinfo {volume} {106}},\
  \bibinfo {pages} {L061305} (\bibinfo {year} {2022})},\ \Eprint
  {https://arxiv.org/abs/2208.14772} {arXiv:2208.14772 [hep-ph]} \BibitemShut
  {NoStop}%
\bibitem [{\citenamefont {Christy}\ and\ \citenamefont
  {Bosted}(2007)}]{Christy:2007ve}%
  \BibitemOpen
  \bibfield  {author} {\bibinfo {author} {\bibfnamefont {M.~E.}\ \bibnamefont
  {Christy}}\ and\ \bibinfo {author} {\bibfnamefont {P.~E.}\ \bibnamefont
  {Bosted}},\ }\bibfield  {title} {\bibinfo {title} {{Empirical Fit to
  Precision Inclusive Electron-Proton Cross Sections in the Resonance
  Region}},\ }\href@noop {} {\  (\bibinfo {year} {2007})},\ \Eprint
  {https://arxiv.org/abs/0712.3731} {arXiv:0712.3731 [hep-ph]} \BibitemShut
  {NoStop}%
\bibitem [{\citenamefont {Bosted}\ and\ \citenamefont
  {Christy}(2008)}]{Bosted:2007xd}%
  \BibitemOpen
  \bibfield  {author} {\bibinfo {author} {\bibfnamefont {P.~E.}\ \bibnamefont
  {Bosted}}\ and\ \bibinfo {author} {\bibfnamefont {M.~E.}\ \bibnamefont
  {Christy}},\ }\bibfield  {title} {\bibinfo {title} {{Empirical fit to
  inelastic electron-deuteron and electron-neutron resonance region transverse
  cross-sections}},\ }\href {https://doi.org/10.1103/PhysRevC.77.065206}
  {\bibfield  {journal} {\bibinfo  {journal} {Phys. Rev. C}\ }\textbf {\bibinfo
  {volume} {77}},\ \bibinfo {pages} {065206} (\bibinfo {year} {2008})},\
  \Eprint {https://arxiv.org/abs/0711.0159} {arXiv:0711.0159 [hep-ph]}
  \BibitemShut {NoStop}%
\bibitem [{\citenamefont {Leitner}\ \emph
  {et~al.}(2006{\natexlab{a}})\citenamefont {Leitner}, \citenamefont
  {Alvarez-Ruso},\ and\ \citenamefont {Mosel}}]{Leitner:2006ww}%
  \BibitemOpen
  \bibfield  {author} {\bibinfo {author} {\bibfnamefont {T.}~\bibnamefont
  {Leitner}}, \bibinfo {author} {\bibfnamefont {L.}~\bibnamefont
  {Alvarez-Ruso}},\ and\ \bibinfo {author} {\bibfnamefont {U.}~\bibnamefont
  {Mosel}},\ }\bibfield  {title} {\bibinfo {title} {{Charged current neutrino
  nucleus interactions at intermediate energies}},\ }\href
  {https://doi.org/10.1103/PhysRevC.73.065502} {\bibfield  {journal} {\bibinfo
  {journal} {Phys. Rev.}\ }\textbf {\bibinfo {volume} {C73}},\ \bibinfo {pages}
  {065502} (\bibinfo {year} {2006}{\natexlab{a}})},\ \Eprint
  {https://arxiv.org/abs/nucl-th/0601103} {arXiv:nucl-th/0601103} \BibitemShut
  {NoStop}%
\bibitem [{\citenamefont {Lalakulich}\ \emph {et~al.}(2010)\citenamefont
  {Lalakulich}, \citenamefont {Leitner}, \citenamefont {Buss},\ and\
  \citenamefont {Mosel}}]{Lalakulich:2010ss}%
  \BibitemOpen
  \bibfield  {author} {\bibinfo {author} {\bibfnamefont {O.}~\bibnamefont
  {Lalakulich}}, \bibinfo {author} {\bibfnamefont {T.}~\bibnamefont {Leitner}},
  \bibinfo {author} {\bibfnamefont {O.}~\bibnamefont {Buss}},\ and\ \bibinfo
  {author} {\bibfnamefont {U.}~\bibnamefont {Mosel}},\ }\bibfield  {title}
  {\bibinfo {title} {{One pion production in neutrino reactions: including
  non-resonant background}},\ }\href
  {https://doi.org/10.1103/PhysRevD.82.093001} {\bibfield  {journal} {\bibinfo
  {journal} {Phys.Rev.}\ }\textbf {\bibinfo {volume} {D82}},\ \bibinfo {pages}
  {093001} (\bibinfo {year} {2010})},\ \Eprint
  {https://arxiv.org/abs/1007.0925} {arXiv:1007.0925 [hep-ph]} \BibitemShut
  {NoStop}%
\bibitem [{\citenamefont {Lalakulich}\ \emph {et~al.}(2006)\citenamefont
  {Lalakulich}, \citenamefont {Paschos},\ and\ \citenamefont
  {Piranishvili}}]{Lalakulich:2006sw}%
  \BibitemOpen
  \bibfield  {author} {\bibinfo {author} {\bibfnamefont {O.}~\bibnamefont
  {Lalakulich}}, \bibinfo {author} {\bibfnamefont {E.~A.}\ \bibnamefont
  {Paschos}},\ and\ \bibinfo {author} {\bibfnamefont {G.}~\bibnamefont
  {Piranishvili}},\ }\bibfield  {title} {\bibinfo {title} {Resonance production
  by neutrinos: The second resonance region},\ }\href@noop {} {\bibfield
  {journal} {\bibinfo  {journal} {Phys. Rev.}\ }\textbf {\bibinfo {volume}
  {D74}},\ \bibinfo {pages} {014009} (\bibinfo {year} {2006})},\ \Eprint
  {https://arxiv.org/abs/hep-ph/0602210} {hep-ph/0602210} \BibitemShut
  {NoStop}%
\bibitem [{\citenamefont {Walecka}(1975)}]{Walecka:1975}%
  \BibitemOpen
  \bibfield  {author} {\bibinfo {author} {\bibfnamefont {D.}~\bibnamefont
  {Walecka}},\ }\bibinfo {title} {Semileptonic weak interactions in nuclei},\
  in\ \href@noop {} {\emph {\bibinfo {booktitle} {Muon Physics}}},\ \bibinfo
  {editor} {edited by\ \bibinfo {editor} {\bibfnamefont {V.}~\bibnamefont
  {Hughes}}\ and\ \bibinfo {editor} {\bibfnamefont {C.}~\bibnamefont {Wu}}}\
  (\bibinfo  {publisher} {Academic Press, New York},\ \bibinfo {year}
  {1975})\BibitemShut {NoStop}%
\bibitem [{\citenamefont {Sjostrand}\ \emph {et~al.}(2006)\citenamefont
  {Sjostrand}, \citenamefont {Mrenna},\ and\ \citenamefont
  {Skands}}]{Sjostrand:2006za}%
  \BibitemOpen
  \bibfield  {author} {\bibinfo {author} {\bibfnamefont {T.}~\bibnamefont
  {Sjostrand}}, \bibinfo {author} {\bibfnamefont {S.}~\bibnamefont {Mrenna}},\
  and\ \bibinfo {author} {\bibfnamefont {P.~Z.}\ \bibnamefont {Skands}},\
  }\bibfield  {title} {\bibinfo {title} {{PYTHIA 6.4 Physics and Manual}},\
  }\href {https://doi.org/10.1088/1126-6708/2006/05/026} {\bibfield  {journal}
  {\bibinfo  {journal} {JHEP}\ }\textbf {\bibinfo {volume} {05}},\ \bibinfo
  {pages} {026}},\ \Eprint {https://arxiv.org/abs/hep-ph/0603175}
  {arXiv:hep-ph/0603175} \BibitemShut {NoStop}%
\bibitem [{\citenamefont {Lalakulich}\ \emph {et~al.}(2012)\citenamefont
  {Lalakulich}, \citenamefont {Gallmeister},\ and\ \citenamefont
  {Mosel}}]{Lalakulich:2012gm}%
  \BibitemOpen
  \bibfield  {author} {\bibinfo {author} {\bibfnamefont {O.}~\bibnamefont
  {Lalakulich}}, \bibinfo {author} {\bibfnamefont {K.}~\bibnamefont
  {Gallmeister}},\ and\ \bibinfo {author} {\bibfnamefont {U.}~\bibnamefont
  {Mosel}},\ }\bibfield  {title} {\bibinfo {title} {{Neutrino and antineutrino
  induced reactions with nuclei between 1 and 50 GeV}},\ }\href@noop {}
  {\bibfield  {journal} {\bibinfo  {journal} {Phys.Rev.}\ }\textbf {\bibinfo
  {volume} {C86}},\ \bibinfo {pages} {014607} (\bibinfo {year} {2012})},\
  \Eprint {https://arxiv.org/abs/1205.1061} {arXiv:1205.1061 [nucl-th]}
  \BibitemShut {NoStop}%
\bibitem [{\citenamefont {Alberico}\ \emph {et~al.}(1998)\citenamefont
  {Alberico}, \citenamefont {Chanfray}, \citenamefont {Delorme}, \citenamefont
  {Ericson},\ and\ \citenamefont {Molinari}}]{Alberico:1997jg}%
  \BibitemOpen
  \bibfield  {author} {\bibinfo {author} {\bibfnamefont {W.~M.}\ \bibnamefont
  {Alberico}}, \bibinfo {author} {\bibfnamefont {G.}~\bibnamefont {Chanfray}},
  \bibinfo {author} {\bibfnamefont {J.}~\bibnamefont {Delorme}}, \bibinfo
  {author} {\bibfnamefont {M.}~\bibnamefont {Ericson}},\ and\ \bibinfo {author}
  {\bibfnamefont {A.}~\bibnamefont {Molinari}},\ }\bibfield  {title} {\bibinfo
  {title} {{The semiclassical approach to the exclusive electron scattering}},\
  }\href {https://doi.org/10.1016/S0375-9474(98)00160-2} {\bibfield  {journal}
  {\bibinfo  {journal} {Nucl. Phys.}\ }\textbf {\bibinfo {volume} {A634}},\
  \bibinfo {pages} {233} (\bibinfo {year} {1998})},\ \Eprint
  {https://arxiv.org/abs/nucl-th/9708043} {arXiv:nucl-th/9708043 [nucl-th]}
  \BibitemShut {NoStop}%
\bibitem [{\citenamefont {Sobczyk}\ and\ \citenamefont
  {Bacca}(2023)}]{Sobczyk:2023mey}%
  \BibitemOpen
  \bibfield  {author} {\bibinfo {author} {\bibfnamefont {J.~E.}\ \bibnamefont
  {Sobczyk}}\ and\ \bibinfo {author} {\bibfnamefont {S.}~\bibnamefont
  {Bacca}},\ }\bibfield  {title} {\bibinfo {title} {{$^{16}$O spectral function
  from coupled-cluster theory: applications to lepton-nucleus scattering}},\
  }\href@noop {} {\  (\bibinfo {year} {2023})},\ \Eprint
  {https://arxiv.org/abs/2309.00355} {arXiv:2309.00355 [nucl-th]} \BibitemShut
  {NoStop}%
\bibitem [{\citenamefont {Gale}\ \emph {et~al.}(1990)\citenamefont {Gale},
  \citenamefont {Welke}, \citenamefont {Prakash}, \citenamefont {Lee},\ and\
  \citenamefont {Das~Gupta}}]{Gale:1989dm}%
  \BibitemOpen
  \bibfield  {author} {\bibinfo {author} {\bibfnamefont {C.}~\bibnamefont
  {Gale}}, \bibinfo {author} {\bibfnamefont {G.}~\bibnamefont {Welke}},
  \bibinfo {author} {\bibfnamefont {M.}~\bibnamefont {Prakash}}, \bibinfo
  {author} {\bibfnamefont {S.}~\bibnamefont {Lee}},\ and\ \bibinfo {author}
  {\bibfnamefont {S.}~\bibnamefont {Das~Gupta}},\ }\bibfield  {title} {\bibinfo
  {title} {{Transverse momenta, nuclear equation of state, and
  momentum-dependent interactions in heavy-ion collisions}},\ }\href
  {https://doi.org/10.1103/PhysRevC.41.1545} {\bibfield  {journal} {\bibinfo
  {journal} {Phys.Rev.}\ }\textbf {\bibinfo {volume} {C41}},\ \bibinfo {pages}
  {1545} (\bibinfo {year} {1990})}\BibitemShut {NoStop}%
\bibitem [{\citenamefont {Cooper}\ \emph {et~al.}(1993)\citenamefont {Cooper},
  \citenamefont {Hama}, \citenamefont {Clark},\ and\ \citenamefont
  {Mercer}}]{Cooper:1993nx}%
  \BibitemOpen
  \bibfield  {author} {\bibinfo {author} {\bibfnamefont {E.~D.}\ \bibnamefont
  {Cooper}}, \bibinfo {author} {\bibfnamefont {S.}~\bibnamefont {Hama}},
  \bibinfo {author} {\bibfnamefont {B.~C.}\ \bibnamefont {Clark}},\ and\
  \bibinfo {author} {\bibfnamefont {R.~L.}\ \bibnamefont {Mercer}},\ }\bibfield
   {title} {\bibinfo {title} {{Global Dirac phenomenology for proton nucleus
  elastic scattering}},\ }\href {https://doi.org/10.1103/PhysRevC.47.297}
  {\bibfield  {journal} {\bibinfo  {journal} {Phys. Rev.}\ }\textbf {\bibinfo
  {volume} {C47}},\ \bibinfo {pages} {297} (\bibinfo {year}
  {1993})}\BibitemShut {NoStop}%
\bibitem [{\citenamefont {Liang}\ \emph {et~al.}(2022)\citenamefont {Liang}
  \emph {et~al.}}]{JeffersonLabHallCE94-110:2004nsn}%
  \BibitemOpen
  \bibfield  {author} {\bibinfo {author} {\bibfnamefont {Y.}~\bibnamefont
  {Liang}} \emph {et~al.} (\bibinfo {collaboration} {Jefferson Lab Hall C
  E94-110}),\ }\bibfield  {title} {\bibinfo {title} {{Measurement of
  R=\ensuremath{\sigma}L/\ensuremath{\sigma}T and the separated longitudinal
  and transverse structure functions in the nucleon-resonance region}},\ }\href
  {https://doi.org/10.1103/PhysRevC.105.065205} {\bibfield  {journal} {\bibinfo
   {journal} {Phys. Rev. C}\ }\textbf {\bibinfo {volume} {105}},\ \bibinfo
  {pages} {065205} (\bibinfo {year} {2022})},\ \Eprint
  {https://arxiv.org/abs/nucl-ex/0410027} {arXiv:nucl-ex/0410027} \BibitemShut
  {NoStop}%
\bibitem [{{JLABdata}()}]{JLABdata_E94-110}%
  \BibitemOpen
  {JLABdata},\ \href@noop {} {}\bibinfo {note}
  {\url{https://hallcweb.jlab.org/resdata/database/cs_94_sys.dat}}\BibitemShut
  {NoStop}%
\bibitem [{\citenamefont {Maieron}\ \emph {et~al.}(2009)\citenamefont
  {Maieron}, \citenamefont {Amaro}, \citenamefont {Barbaro}, \citenamefont
  {Caballero}, \citenamefont {Donnelly},\ and\ \citenamefont
  {Williamson}}]{Maieron:2009an}%
  \BibitemOpen
  \bibfield  {author} {\bibinfo {author} {\bibfnamefont {C.}~\bibnamefont
  {Maieron}}, \bibinfo {author} {\bibfnamefont {J.~E.}\ \bibnamefont {Amaro}},
  \bibinfo {author} {\bibfnamefont {M.~B.}\ \bibnamefont {Barbaro}}, \bibinfo
  {author} {\bibfnamefont {J.~A.}\ \bibnamefont {Caballero}}, \bibinfo {author}
  {\bibfnamefont {T.~W.}\ \bibnamefont {Donnelly}},\ and\ \bibinfo {author}
  {\bibfnamefont {C.~F.}\ \bibnamefont {Williamson}},\ }\bibfield  {title}
  {\bibinfo {title} {{Superscaling of non-quasielastic electron-nucleus
  scattering}},\ }\href {https://doi.org/10.1103/PhysRevC.80.035504} {\bibfield
   {journal} {\bibinfo  {journal} {Phys. Rev. C}\ }\textbf {\bibinfo {volume}
  {80}},\ \bibinfo {pages} {035504} (\bibinfo {year} {2009})},\ \Eprint
  {https://arxiv.org/abs/0907.1841} {arXiv:0907.1841 [nucl-th]} \BibitemShut
  {NoStop}%
\bibitem [{\citenamefont {Megias}\ \emph {et~al.}(2016)\citenamefont {Megias},
  \citenamefont {Amaro}, \citenamefont {Barbaro}, \citenamefont {Caballero},\
  and\ \citenamefont {Donnelly}}]{Megias:2016lke}%
  \BibitemOpen
  \bibfield  {author} {\bibinfo {author} {\bibfnamefont {G.~D.}\ \bibnamefont
  {Megias}}, \bibinfo {author} {\bibfnamefont {J.~E.}\ \bibnamefont {Amaro}},
  \bibinfo {author} {\bibfnamefont {M.~B.}\ \bibnamefont {Barbaro}}, \bibinfo
  {author} {\bibfnamefont {J.~A.}\ \bibnamefont {Caballero}},\ and\ \bibinfo
  {author} {\bibfnamefont {T.~W.}\ \bibnamefont {Donnelly}},\ }\bibfield
  {title} {\bibinfo {title} {{Inclusive electron scattering within the SuSAv2
  meson-exchange current approach}},\ }\href
  {https://doi.org/10.1103/PhysRevD.94.013012} {\bibfield  {journal} {\bibinfo
  {journal} {Phys. Rev.}\ }\textbf {\bibinfo {volume} {D94}},\ \bibinfo {pages}
  {013012} (\bibinfo {year} {2016})},\ \Eprint
  {https://arxiv.org/abs/1603.08396} {arXiv:1603.08396 [nucl-th]} \BibitemShut
  {NoStop}%
\bibitem [{\citenamefont {Gonzalez-Rosa}\ \emph {et~al.}(2022)\citenamefont
  {Gonzalez-Rosa}, \citenamefont {Megias}, \citenamefont {Caballero},\ and\
  \citenamefont {Barbaro}}]{Gonzalez-Rosa:2022ltp}%
  \BibitemOpen
  \bibfield  {author} {\bibinfo {author} {\bibfnamefont {J.}~\bibnamefont
  {Gonzalez-Rosa}}, \bibinfo {author} {\bibfnamefont {G.~D.}\ \bibnamefont
  {Megias}}, \bibinfo {author} {\bibfnamefont {J.~A.}\ \bibnamefont
  {Caballero}},\ and\ \bibinfo {author} {\bibfnamefont {M.~B.}\ \bibnamefont
  {Barbaro}},\ }\bibfield  {title} {\bibinfo {title} {{SuSAv2 model for
  inelastic neutrino-nucleus scattering}},\ }\href
  {https://doi.org/10.1103/PhysRevD.105.093009} {\bibfield  {journal} {\bibinfo
   {journal} {Phys. Rev. D}\ }\textbf {\bibinfo {volume} {105}},\ \bibinfo
  {pages} {093009} (\bibinfo {year} {2022})},\ \Eprint
  {https://arxiv.org/abs/2203.12308} {arXiv:2203.12308 [nucl-th]} \BibitemShut
  {NoStop}%
\bibitem [{\citenamefont {E.~Byckling}(1973)}]{Byck:1973}%
  \BibitemOpen
  \bibfield  {author} {\bibinfo {author} {\bibfnamefont {K.~K.}\ \bibnamefont
  {E.~Byckling}},\ }\href@noop {} {\emph {\bibinfo {title} {Particle
  Kinematics}}}\ (\bibinfo  {publisher} {John Wiley and Sons},\ \bibinfo
  {address} {London},\ \bibinfo {year} {1973})\BibitemShut {NoStop}%
\bibitem [{\citenamefont {Benhar}\ \emph {et~al.}(2008)\citenamefont {Benhar},
  \citenamefont {Day},\ and\ \citenamefont {Sick}}]{Benhar:2006wy}%
  \BibitemOpen
  \bibfield  {author} {\bibinfo {author} {\bibfnamefont {O.}~\bibnamefont
  {Benhar}}, \bibinfo {author} {\bibfnamefont {D.}~\bibnamefont {Day}},\ and\
  \bibinfo {author} {\bibfnamefont {I.}~\bibnamefont {Sick}},\ }\bibfield
  {title} {\bibinfo {title} {{Inclusive quasi-elastic electron-nucleus
  scattering}},\ }\href {https://doi.org/10.1103/RevModPhys.80.189} {\bibfield
  {journal} {\bibinfo  {journal} {Rev. Mod. Phys.}\ }\textbf {\bibinfo {volume}
  {80}},\ \bibinfo {pages} {189} (\bibinfo {year} {2008})},\ \Eprint
  {https://arxiv.org/abs/nucl-ex/0603029} {arXiv:nucl-ex/0603029} \BibitemShut
  {NoStop}%
\bibitem [{\citenamefont {Formaggio}\ and\ \citenamefont
  {Zeller}(2012)}]{Formaggio:2012cpf}%
  \BibitemOpen
  \bibfield  {author} {\bibinfo {author} {\bibfnamefont {J.~A.}\ \bibnamefont
  {Formaggio}}\ and\ \bibinfo {author} {\bibfnamefont {G.~P.}\ \bibnamefont
  {Zeller}},\ }\bibfield  {title} {\bibinfo {title} {{From eV to EeV: Neutrino
  Cross Sections Across Energy Scales}},\ }\href
  {https://doi.org/10.1103/RevModPhys.84.1307} {\bibfield  {journal} {\bibinfo
  {journal} {Rev. Mod. Phys.}\ }\textbf {\bibinfo {volume} {84}},\ \bibinfo
  {pages} {1307} (\bibinfo {year} {2012})},\ \Eprint
  {https://arxiv.org/abs/1305.7513} {arXiv:1305.7513 [hep-ex]} \BibitemShut
  {NoStop}%
\bibitem [{\citenamefont {Leitner}\ \emph
  {et~al.}(2006{\natexlab{b}})\citenamefont {Leitner}, \citenamefont
  {Alvarez-Ruso},\ and\ \citenamefont {Mosel}}]{Leitner:2006sp}%
  \BibitemOpen
  \bibfield  {author} {\bibinfo {author} {\bibfnamefont {T.}~\bibnamefont
  {Leitner}}, \bibinfo {author} {\bibfnamefont {L.}~\bibnamefont
  {Alvarez-Ruso}},\ and\ \bibinfo {author} {\bibfnamefont {U.}~\bibnamefont
  {Mosel}},\ }\bibfield  {title} {\bibinfo {title} {{Neutral current neutrino
  nucleus interactions at intermediate energies}},\ }\href
  {https://doi.org/10.1103/PhysRevC.74.065502} {\bibfield  {journal} {\bibinfo
  {journal} {Phys. Rev.}\ }\textbf {\bibinfo {volume} {C74}},\ \bibinfo {pages}
  {065502} (\bibinfo {year} {2006}{\natexlab{b}})},\ \Eprint
  {https://arxiv.org/abs/nucl-th/0606058} {arXiv:nucl-th/0606058} \BibitemShut
  {NoStop}%
\bibitem [{\citenamefont {Martinez-Consentino}\ \emph
  {et~al.}(2021)\citenamefont {Martinez-Consentino}, \citenamefont {Amaro},\
  and\ \citenamefont {Ruiz~Simo}}]{Martinez-Consentino:2021vcs}%
  \BibitemOpen
  \bibfield  {author} {\bibinfo {author} {\bibfnamefont {V.~L.}\ \bibnamefont
  {Martinez-Consentino}}, \bibinfo {author} {\bibfnamefont {J.~E.}\
  \bibnamefont {Amaro}},\ and\ \bibinfo {author} {\bibfnamefont
  {I.}~\bibnamefont {Ruiz~Simo}},\ }\bibfield  {title} {\bibinfo {title}
  {{Semiempirical formula for electroweak response functions in the two-nucleon
  emission channel in neutrino-nucleus scattering}},\ }\href
  {https://doi.org/10.1103/PhysRevD.104.113006} {\bibfield  {journal} {\bibinfo
   {journal} {Phys. Rev. D}\ }\textbf {\bibinfo {volume} {104}},\ \bibinfo
  {pages} {113006} (\bibinfo {year} {2021})},\ \Eprint
  {https://arxiv.org/abs/2109.00854} {arXiv:2109.00854 [nucl-th]} \BibitemShut
  {NoStop}%
\bibitem [{\citenamefont {Bodek}\ and\ \citenamefont
  {Yang}(2010)}]{Bodek:2010km}%
  \BibitemOpen
  \bibfield  {author} {\bibinfo {author} {\bibfnamefont {A.}~\bibnamefont
  {Bodek}}\ and\ \bibinfo {author} {\bibfnamefont {U.-k.}\ \bibnamefont
  {Yang}},\ }\bibfield  {title} {\bibinfo {title} {{Axial and Vector Structure
  Functions for Electron- and Neutrino- Nucleon Scattering Cross Sections at
  all $Q^2$ using Effective Leading order Parton Distribution Functions}},\
  }\href@noop {} {\  (\bibinfo {year} {2010})},\ \Eprint
  {https://arxiv.org/abs/1011.6592} {arXiv:1011.6592 [hep-ph]} \BibitemShut
  {NoStop}%
\bibitem [{\citenamefont {O'~Connell}\ \emph {et~al.}(1972)\citenamefont
  {O'~Connell}, \citenamefont {Donnelly},\ and\ \citenamefont
  {Walecka}}]{OConnell:1972edu}%
  \BibitemOpen
  \bibfield  {author} {\bibinfo {author} {\bibfnamefont {J.~S.}\ \bibnamefont
  {O'~Connell}}, \bibinfo {author} {\bibfnamefont {T.~W.}\ \bibnamefont
  {Donnelly}},\ and\ \bibinfo {author} {\bibfnamefont {J.~D.}\ \bibnamefont
  {Walecka}},\ }\bibfield  {title} {\bibinfo {title} {{Semileptonic weak
  interactions with $C^{12}$}},\ }\href
  {https://doi.org/10.1103/PhysRevC.6.719} {\bibfield  {journal} {\bibinfo
  {journal} {Phys. Rev.}\ }\textbf {\bibinfo {volume} {C6}},\ \bibinfo {pages}
  {719} (\bibinfo {year} {1972})}\BibitemShut {NoStop}%
\bibitem [{\citenamefont {Benhar}\ \emph {et~al.}(2006)\citenamefont {Benhar},
  \citenamefont {Day},\ and\ \citenamefont {Sick}}]{Benhar:2006er}%
  \BibitemOpen
  \bibfield  {author} {\bibinfo {author} {\bibfnamefont {O.}~\bibnamefont
  {Benhar}}, \bibinfo {author} {\bibfnamefont {D.}~\bibnamefont {Day}},\ and\
  \bibinfo {author} {\bibfnamefont {I.}~\bibnamefont {Sick}},\ }\bibfield
  {title} {\bibinfo {title} {{An Archive for quasi-elastic electron-nucleus
  scattering data}},\ }\href@noop {} {\  (\bibinfo {year} {2006})},\ \Eprint
  {https://arxiv.org/abs/nucl-ex/0603032} {arXiv:nucl-ex/0603032 [nucl-ex]}
  \BibitemShut {NoStop}%
\bibitem [{\citenamefont {Kolbe}\ \emph {et~al.}(1994)\citenamefont {Kolbe},
  \citenamefont {Langanke},\ and\ \citenamefont {Krewald}}]{Kolbe:1994xb}%
  \BibitemOpen
  \bibfield  {author} {\bibinfo {author} {\bibfnamefont {E.}~\bibnamefont
  {Kolbe}}, \bibinfo {author} {\bibfnamefont {K.}~\bibnamefont {Langanke}},\
  and\ \bibinfo {author} {\bibfnamefont {S.}~\bibnamefont {Krewald}},\
  }\bibfield  {title} {\bibinfo {title} {{Neutrino induced reactions on C-12
  within the continuum random phase approximation}},\ }\href
  {https://doi.org/10.1103/PhysRevC.49.1122} {\bibfield  {journal} {\bibinfo
  {journal} {Phys. Rev. C}\ }\textbf {\bibinfo {volume} {49}},\ \bibinfo
  {pages} {1122} (\bibinfo {year} {1994})}\BibitemShut {NoStop}%
\bibitem [{\citenamefont {Kolbe}\ \emph {et~al.}(1995)\citenamefont {Kolbe},
  \citenamefont {Langanke}, \citenamefont {Thielemann},\ and\ \citenamefont
  {Vogel}}]{Kolbe:1995af}%
  \BibitemOpen
  \bibfield  {author} {\bibinfo {author} {\bibfnamefont {E.}~\bibnamefont
  {Kolbe}}, \bibinfo {author} {\bibfnamefont {K.}~\bibnamefont {Langanke}},
  \bibinfo {author} {\bibfnamefont {F.-K.}\ \bibnamefont {Thielemann}},\ and\
  \bibinfo {author} {\bibfnamefont {P.}~\bibnamefont {Vogel}},\ }\bibfield
  {title} {\bibinfo {title} {{Inclusive C-12(nu/mu,mu)N-12 reaction in the
  continuum random phase approximation}},\ }\href
  {https://doi.org/10.1103/PhysRevC.52.3437} {\bibfield  {journal} {\bibinfo
  {journal} {Phys. Rev.}\ }\textbf {\bibinfo {volume} {C52}},\ \bibinfo {pages}
  {3437} (\bibinfo {year} {1995})}\BibitemShut {NoStop}%
\bibitem [{\citenamefont {Pandey}\ \emph {et~al.}(2015)\citenamefont {Pandey},
  \citenamefont {Jachowicz}, \citenamefont {Van~Cuyck}, \citenamefont
  {Ryckebusch},\ and\ \citenamefont {Martini}}]{Pandey:2014tza}%
  \BibitemOpen
  \bibfield  {author} {\bibinfo {author} {\bibfnamefont {V.}~\bibnamefont
  {Pandey}}, \bibinfo {author} {\bibfnamefont {N.}~\bibnamefont {Jachowicz}},
  \bibinfo {author} {\bibfnamefont {T.}~\bibnamefont {Van~Cuyck}}, \bibinfo
  {author} {\bibfnamefont {J.}~\bibnamefont {Ryckebusch}},\ and\ \bibinfo
  {author} {\bibfnamefont {M.}~\bibnamefont {Martini}},\ }\bibfield  {title}
  {\bibinfo {title} {{Low-energy excitations and quasielastic contribution to
  electron-nucleus and neutrino-nucleus scattering in the continuum
  random-phase approximation}},\ }\href
  {https://doi.org/10.1103/PhysRevC.92.024606} {\bibfield  {journal} {\bibinfo
  {journal} {Phys. Rev.}\ }\textbf {\bibinfo {volume} {C92}},\ \bibinfo {pages}
  {024606} (\bibinfo {year} {2015})},\ \Eprint
  {https://arxiv.org/abs/1412.4624} {arXiv:1412.4624 [nucl-th]} \BibitemShut
  {NoStop}%
\bibitem [{\citenamefont {Jachowicz}\ \emph {et~al.}(2019)\citenamefont
  {Jachowicz}, \citenamefont {Van~Dessel},\ and\ \citenamefont
  {Nikolakopoulos}}]{Jachowicz:2019eul}%
  \BibitemOpen
  \bibfield  {author} {\bibinfo {author} {\bibfnamefont {N.}~\bibnamefont
  {Jachowicz}}, \bibinfo {author} {\bibfnamefont {N.}~\bibnamefont
  {Van~Dessel}},\ and\ \bibinfo {author} {\bibfnamefont {A.}~\bibnamefont
  {Nikolakopoulos}},\ }\bibfield  {title} {\bibinfo {title} {{Low-energy
  neutrino scattering in experiment and astrophysics}},\ }\href
  {https://doi.org/10.1088/1361-6471/ab25d4} {\bibfield  {journal} {\bibinfo
  {journal} {J. Phys. G}\ }\textbf {\bibinfo {volume} {46}},\ \bibinfo {pages}
  {084003} (\bibinfo {year} {2019})},\ \Eprint
  {https://arxiv.org/abs/1906.08191} {arXiv:1906.08191 [nucl-th]} \BibitemShut
  {NoStop}%
\bibitem [{\citenamefont {Dai}\ \emph {et~al.}(2018)\citenamefont {Dai} \emph
  {et~al.}}]{JeffersonLabHallA:2018zyx}%
  \BibitemOpen
  \bibfield  {author} {\bibinfo {author} {\bibfnamefont {H.}~\bibnamefont
  {Dai}} \emph {et~al.} (\bibinfo {collaboration} {Jefferson Lab Hall A}),\
  }\bibfield  {title} {\bibinfo {title} {{First Measurement of the
  Ti$(e,e^\prime){\rm X}$ Cross Section at Jefferson Lab}},\ }\href
  {https://doi.org/10.1103/PhysRevC.98.014617} {\bibfield  {journal} {\bibinfo
  {journal} {Phys. Rev.}\ }\textbf {\bibinfo {volume} {C98}},\ \bibinfo {pages}
  {014617} (\bibinfo {year} {2018})},\ \Eprint
  {https://arxiv.org/abs/1803.01910} {arXiv:1803.01910 [nucl-ex]} \BibitemShut
  {NoStop}%
\bibitem [{\citenamefont {Mosel}\ and\ \citenamefont
  {Gallmeister}(2019)}]{Mosel:2018qmv}%
  \BibitemOpen
  \bibfield  {author} {\bibinfo {author} {\bibfnamefont {U.}~\bibnamefont
  {Mosel}}\ and\ \bibinfo {author} {\bibfnamefont {K.}~\bibnamefont
  {Gallmeister}},\ }\bibfield  {title} {\bibinfo {title} {{Cross sections for
  A(e,e')X reactions}},\ }\href {https://doi.org/10.1103/PhysRevC.99.064605}
  {\bibfield  {journal} {\bibinfo  {journal} {Phys. Rev.}\ }\textbf {\bibinfo
  {volume} {C99}},\ \bibinfo {pages} {064605} (\bibinfo {year} {2019})},\
  \Eprint {https://arxiv.org/abs/1811.10637} {arXiv:1811.10637 [nucl-ex]}
  \BibitemShut {NoStop}%
\bibitem [{\citenamefont {Papadopoulou}\ \emph {et~al.}(2021)\citenamefont
  {Papadopoulou}, \citenamefont {Ashkenazi}, \citenamefont {Gardiner},
  \citenamefont {Betancourt}, \citenamefont {Dytman}, \citenamefont
  {Weinstein}, \citenamefont {Piasetzky}, \citenamefont {Hauenstein},
  \citenamefont {Khachatryan}, \citenamefont {Dolan}, \citenamefont {Megias},\
  and\ \citenamefont {Hen}}]{PhysRevD.103.113003}%
  \BibitemOpen
  \bibfield  {author} {\bibinfo {author} {\bibfnamefont {A.}~\bibnamefont
  {Papadopoulou}}, \bibinfo {author} {\bibfnamefont {A.}~\bibnamefont
  {Ashkenazi}}, \bibinfo {author} {\bibfnamefont {S.}~\bibnamefont {Gardiner}},
  \bibinfo {author} {\bibfnamefont {M.}~\bibnamefont {Betancourt}}, \bibinfo
  {author} {\bibfnamefont {S.}~\bibnamefont {Dytman}}, \bibinfo {author}
  {\bibfnamefont {L.~B.}\ \bibnamefont {Weinstein}}, \bibinfo {author}
  {\bibfnamefont {E.}~\bibnamefont {Piasetzky}}, \bibinfo {author}
  {\bibfnamefont {F.}~\bibnamefont {Hauenstein}}, \bibinfo {author}
  {\bibfnamefont {M.}~\bibnamefont {Khachatryan}}, \bibinfo {author}
  {\bibfnamefont {S.}~\bibnamefont {Dolan}}, \bibinfo {author} {\bibfnamefont
  {G.~D.}\ \bibnamefont {Megias}},\ and\ \bibinfo {author} {\bibfnamefont
  {O.}~\bibnamefont {Hen}} (\bibinfo {collaboration} {$e4\ensuremath{\nu}$
  Collaboration}),\ }\bibfield  {title} {\bibinfo {title} {Inclusive electron
  scattering and the genie neutrino event generator},\ }\href
  {https://doi.org/10.1103/PhysRevD.103.113003} {\bibfield  {journal} {\bibinfo
   {journal} {Phys. Rev. D}\ }\textbf {\bibinfo {volume} {103}},\ \bibinfo
  {pages} {113003} (\bibinfo {year} {2021})}\BibitemShut {NoStop}%
\bibitem [{\citenamefont {Mosel}\ and\ \citenamefont
  {Gallmeister}(2018)}]{Mosel:2017anp}%
  \BibitemOpen
  \bibfield  {author} {\bibinfo {author} {\bibfnamefont {U.}~\bibnamefont
  {Mosel}}\ and\ \bibinfo {author} {\bibfnamefont {K.}~\bibnamefont
  {Gallmeister}},\ }\bibfield  {title} {\bibinfo {title}
  {{Muon-neutrino-induced charged current cross section without pions:
  Theoretical analysis}},\ }\href {https://doi.org/10.1103/PhysRevC.97.045501}
  {\bibfield  {journal} {\bibinfo  {journal} {Phys. Rev. C}\ }\textbf {\bibinfo
  {volume} {97}},\ \bibinfo {pages} {045501} (\bibinfo {year} {2018})},\
  \Eprint {https://arxiv.org/abs/1712.07134} {arXiv:1712.07134 [hep-ex]}
  \BibitemShut {NoStop}%
\bibitem [{\citenamefont {Acero}\ \emph {et~al.}(2021)\citenamefont {Acero}
  \emph {et~al.}}]{NOvA:2021eqi}%
  \BibitemOpen
  \bibfield  {author} {\bibinfo {author} {\bibfnamefont {M.~A.}\ \bibnamefont
  {Acero}} \emph {et~al.} (\bibinfo {collaboration} {NOvA}),\ }\bibfield
  {title} {\bibinfo {title} {{Measurement of the Double-Differential
  Muon-neutrino Charged-Current Inclusive Cross Section in the NOvA Near
  Detector}},\ }\href@noop {} {\  (\bibinfo {year} {2021})},\ \Eprint
  {https://arxiv.org/abs/2109.12220} {arXiv:2109.12220 [hep-ex]} \BibitemShut
  {NoStop}%
\bibitem [{\citenamefont {Cremonesi}(2020)}]{Cremonesi:2020}%
  \BibitemOpen
  \bibfield  {author} {\bibinfo {author} {\bibfnamefont {L.}~\bibnamefont
  {Cremonesi}},\ }\bibfield  {title} {\bibinfo {title} {{Cross-section
  measurements with NOvA}}\ }\href {https://doi.org/10.5281/zenodo.4155399}
  {10.5281/zenodo.4155399} (\bibinfo {year} {2020})\BibitemShut {NoStop}%
\bibitem [{\citenamefont {Hernandez}\ \emph {et~al.}(2007)\citenamefont
  {Hernandez}, \citenamefont {Nieves},\ and\ \citenamefont
  {Valverde}}]{Hernandez:2007qq}%
  \BibitemOpen
  \bibfield  {author} {\bibinfo {author} {\bibfnamefont {E.}~\bibnamefont
  {Hernandez}}, \bibinfo {author} {\bibfnamefont {J.}~\bibnamefont {Nieves}},\
  and\ \bibinfo {author} {\bibfnamefont {M.}~\bibnamefont {Valverde}},\
  }\bibfield  {title} {\bibinfo {title} {{Weak pion production off the
  nucleon}},\ }\href {https://doi.org/10.1103/PhysRevD.76.033005} {\bibfield
  {journal} {\bibinfo  {journal} {Phys. Rev.}\ }\textbf {\bibinfo {volume}
  {D76}},\ \bibinfo {pages} {033005} (\bibinfo {year} {2007})},\ \Eprint
  {https://arxiv.org/abs/hep-ph/0701149} {arXiv:hep-ph/0701149} \BibitemShut
  {NoStop}%
\bibitem [{\citenamefont {Abratenko}\ \emph
  {et~al.}(2021{\natexlab{b}})\citenamefont {Abratenko} \emph
  {et~al.}}]{MicroBooNE:2021cue}%
  \BibitemOpen
  \bibfield  {author} {\bibinfo {author} {\bibfnamefont {P.}~\bibnamefont
  {Abratenko}} \emph {et~al.} (\bibinfo {collaboration} {MicroBooNE}),\
  }\bibfield  {title} {\bibinfo {title} {{First Measurement of Energy-dependent
  Inclusive Muon Neutrino Charged-Current Cross Sections on Argon with the
  MicroBooNE Detector}},\ }\href@noop {} {\  (\bibinfo {year}
  {2021}{\natexlab{b}})},\ \Eprint {https://arxiv.org/abs/2110.14023}
  {arXiv:2110.14023 [hep-ex]} \BibitemShut {NoStop}%
\bibitem [{\citenamefont {Abratenko}\ \emph
  {et~al.}(2023{\natexlab{a}})\citenamefont {Abratenko} \emph
  {et~al.}}]{MicroBooNE:2023cmw}%
  \BibitemOpen
  \bibfield  {author} {\bibinfo {author} {\bibfnamefont {P.}~\bibnamefont
  {Abratenko}} \emph {et~al.} (\bibinfo {collaboration} {MicroBooNE}),\
  }\bibfield  {title} {\bibinfo {title} {{Multidifferential cross section
  measurements of \ensuremath{\nu}\ensuremath{\mu}-argon quasielasticlike
  reactions with the MicroBooNE detector}},\ }\href
  {https://doi.org/10.1103/PhysRevD.108.053002} {\bibfield  {journal} {\bibinfo
   {journal} {Phys. Rev. D}\ }\textbf {\bibinfo {volume} {108}},\ \bibinfo
  {pages} {053002} (\bibinfo {year} {2023}{\natexlab{a}})},\ \Eprint
  {https://arxiv.org/abs/2301.03700} {arXiv:2301.03700 [hep-ex]} \BibitemShut
  {NoStop}%
\bibitem [{\citenamefont {Abratenko}\ \emph
  {et~al.}(2023{\natexlab{b}})\citenamefont {Abratenko} \emph
  {et~al.}}]{MicroBooNE:2023tzj}%
  \BibitemOpen
  \bibfield  {author} {\bibinfo {author} {\bibfnamefont {P.}~\bibnamefont
  {Abratenko}} \emph {et~al.} (\bibinfo {collaboration} {MicroBooNE}),\
  }\bibfield  {title} {\bibinfo {title} {{First Double-Differential Measurement
  of Kinematic Imbalance in Neutrino Interactions with the MicroBooNE
  Detector}},\ }\href {https://doi.org/10.1103/PhysRevLett.131.101802}
  {\bibfield  {journal} {\bibinfo  {journal} {Phys. Rev. Lett.}\ }\textbf
  {\bibinfo {volume} {131}},\ \bibinfo {pages} {101802} (\bibinfo {year}
  {2023}{\natexlab{b}})},\ \Eprint {https://arxiv.org/abs/2301.03706}
  {arXiv:2301.03706 [hep-ex]} \BibitemShut {NoStop}%
\bibitem [{\citenamefont {Lalakulich}\ \emph {et~al.}(2009)\citenamefont
  {Lalakulich}, \citenamefont {Praet}, \citenamefont {Jachowicz}, \citenamefont
  {Ryckebusch}, \citenamefont {Leitner}, \citenamefont {Buss},\ and\
  \citenamefont {Mosel}}]{Lalakulich:2009zza}%
  \BibitemOpen
  \bibfield  {author} {\bibinfo {author} {\bibfnamefont {O.}~\bibnamefont
  {Lalakulich}}, \bibinfo {author} {\bibfnamefont {C.}~\bibnamefont {Praet}},
  \bibinfo {author} {\bibfnamefont {N.}~\bibnamefont {Jachowicz}}, \bibinfo
  {author} {\bibfnamefont {J.}~\bibnamefont {Ryckebusch}}, \bibinfo {author}
  {\bibfnamefont {T.}~\bibnamefont {Leitner}}, \bibinfo {author} {\bibfnamefont
  {O.}~\bibnamefont {Buss}},\ and\ \bibinfo {author} {\bibfnamefont
  {U.}~\bibnamefont {Mosel}},\ }\bibfield  {title} {\bibinfo {title}
  {{Neutrinos and duality}},\ }\href {https://doi.org/10.1063/1.3274170}
  {\bibfield  {journal} {\bibinfo  {journal} {AIP Conf. Proc.}\ }\textbf
  {\bibinfo {volume} {1189}},\ \bibinfo {pages} {276} (\bibinfo {year}
  {2009})}\BibitemShut {NoStop}%
\bibitem [{\citenamefont {Khachatryan}\ \emph {et~al.}(2021)\citenamefont
  {Khachatryan} \emph {et~al.}}]{CLAS:2021neh}%
  \BibitemOpen
  \bibfield  {author} {\bibinfo {author} {\bibfnamefont {M.}~\bibnamefont
  {Khachatryan}} \emph {et~al.} (\bibinfo {collaboration} {CLAS, e4v}),\
  }\bibfield  {title} {\bibinfo {title} {{Electron-beam energy reconstruction
  for neutrino oscillation measurements}},\ }\href
  {https://doi.org/10.1038/s41586-021-04046-5} {\bibfield  {journal} {\bibinfo
  {journal} {Nature}\ }\textbf {\bibinfo {volume} {599}},\ \bibinfo {pages}
  {565} (\bibinfo {year} {2021})}\BibitemShut {NoStop}%
\bibitem [{\citenamefont {Lalakulich}\ \emph {et~al.}(2007)\citenamefont
  {Lalakulich} \emph {et~al.}}]{Lalakulich:2007zz}%
  \BibitemOpen
  \bibfield  {author} {\bibinfo {author} {\bibfnamefont {O.}~\bibnamefont
  {Lalakulich}} \emph {et~al.},\ }\bibfield  {title} {\bibinfo {title}
  {{Duality in neutrino reactions}},\ }\href
  {https://doi.org/10.1063/1.2834485} {\bibfield  {journal} {\bibinfo
  {journal} {AIP Conf. Proc.}\ }\textbf {\bibinfo {volume} {967}},\ \bibinfo
  {pages} {243} (\bibinfo {year} {2007})}\BibitemShut {NoStop}%
\bibitem [{\citenamefont {Christy}(2023)}]{Christy:2023}%
  \BibitemOpen
  \bibfield  {author} {\bibinfo {author} {\bibfnamefont {E.}~\bibnamefont
  {Christy}},\ }\href@noop {} {}\bibinfo {howpublished} {private communication}
  (\bibinfo {year} {2023})\BibitemShut {NoStop}%
\bibitem [{\citenamefont {Mosel}\ and\ \citenamefont
  {Gallmeister}(2016)}]{Mosel:2016uge}%
  \BibitemOpen
  \bibfield  {author} {\bibinfo {author} {\bibfnamefont {U.}~\bibnamefont
  {Mosel}}\ and\ \bibinfo {author} {\bibfnamefont {K.}~\bibnamefont
  {Gallmeister}},\ }\bibfield  {title} {\bibinfo {title} {{Mass dependence and
  isospin dependence of short-range correlated pairs}},\ }\href
  {https://doi.org/10.1103/PhysRevC.94.034610} {\bibfield  {journal} {\bibinfo
  {journal} {Phys. Rev.}\ }\textbf {\bibinfo {volume} {C94}},\ \bibinfo {pages}
  {034610} (\bibinfo {year} {2016})},\ \Eprint
  {https://arxiv.org/abs/1606.06499} {arXiv:1606.06499 [nucl-th]} \BibitemShut
  {NoStop}%
\end{thebibliography}%

\end{document}